\documentclass[a4paper,twocolumn,10pt,accepted=2024-11-16]{quantumarticle}
\pdfoutput=1
\usepackage[utf8]{inputenc}
\usepackage[english]{babel}
\usepackage[T1]{fontenc}
\usepackage{hyperref}

\usepackage{lipsum}
\usepackage[square, numbers,sort&compress]{natbib}

\usepackage{graphicx}
\usepackage{dcolumn}
\usepackage{bm}

\usepackage{multirow}

\usepackage{amsmath,amssymb,amsfonts}
\usepackage{graphicx}
\usepackage{color}
\usepackage{appendix}
\usepackage{amsthm}
\usepackage{tikz}
\usepackage{mleftright}
\usepackage{url}
\usepackage[most]{tcolorbox}

\usepackage{tikz}
\usetikzlibrary{shapes,snakes,positioning,automata,arrows.meta,calc,decorations.markings,math,arrows.meta}
\tikzstyle{vertex}=[circle, draw, inner sep=0pt, minimum size=6pt]
\usepackage{xcolor}
\usepackage{subfig}

\usepackage[export]{adjustbox}

\usepackage{geometry}
\newtheorem{definition}{Definition}

\newcommand{\nn}{\nonumber}

\def\a{\alpha}

\def\t{\tau}

\def\p{\psi}

\def\P{\Psi}

\def\<{\langle}
\def\>{\rangle}

\def\ha{{\hat{a}}}
\def\hb{{\hat{b}}}

\def\t2{{\tilde{1}}}

\usepackage[normalem]{ulem}

\newcommand\encircle[1]{%
  \tikz[baseline=(X.base)] 
    \node (X) [draw, shape=circle, inner sep=0] {\strut #1};}

\begin{document}

	\title{Heralded Optical Entanglement Generation via the Graph Picture of Linear Quantum Networks}

	\author{Seungbeom Chin}
	\email{sbthesy@gmail.com}
	\affiliation{Okinawa Institute of Science and Technology Graduate University, Okinawa 904-0495, Japan \\ International Centre for Theory of Quantum Technologies, University of Gda\'{n}sk, 80-308, Gda\'{n}sk, Poland \\ Department of Electrical and Computer Engineering, Sungkyunkwan University, Suwon 16419, Korea}

	\author{Marcin Karczewski}
	\affiliation{Institute of Spintronics and Quantum Information, Faculty of Physics and Astronomy, Adam Mickiewicz University,  Poland\\International Centre for Theory of Quantum Technologies, University of Gda\'{n}sk, 80-308, Gda\'{n}sk, Poland}

 	\author{Yong-Su Kim}
  \email{yong-su.kim@kist.re.kr}
	\affiliation{Center for Quantum Information, Korea Institute of Science and Technology (KIST), Seoul, 02792, Korea \\
		Division of Quantum Information Technology, KIST School, Korea University of Science and Technology, Seoul 02792, Korea}	
	
	
	\begin{abstract}
Non-destructive heralded entanglement with photons is a valuable resource for quantum information processing. However, they generally entail ancillary particles and modes that amplify the circuit intricacy. To address this challenge, a recent work (\href{https://www.nature.com/articles/s41534-024-00845-6}{npj Quantum Information 10, 67 (2024)}) introduced a graph approach for creating multipartite entanglements with boson subtractions. Nonetheless, it remains an essential intermediate step toward practical heralded schemes: the proposition of heralded subtraction operators in bosonic linear quantum networks. This research establishes comprehensive translation rules from subtraction operators to linear optical operators, which provides a seamless path to design heralded schemes with single photons. Our method begets enhanced or previously unreported schemes for the $N$-partite GHZ state with $2N$ photons, $N$-partite W state with $2N+1$ photons and superposition of $N=3$ GHZ and W states with 9 photons.
Our streamlined approach can straightforwardly design heralded schemes for multipartite entangled states by assembling the operators according to the guidence of sculpting bigraphs, hence significantly simplifies the quantum circuit design process.
 
	\end{abstract}
	
	
	\maketitle

\section{Introduction} 

Quantum entanglement is a fundamental phenomenon in quantum mechanics with profound implications in the fields of quantum information and computation science.
It is an essential resource for several quantum tasks such as quantum computing~\cite{plenio2007introduction,preskill2012quantum} and quantum cryptography~\cite{gisin2002quantum}.

In the domain of optics,  promising approaches for generating entanglement are probabilistic methods leveraging the indistinguishability of photons. The probabilistic entanglement generation methods can be categorized into two distinct types: postselected and heralded schemes.
Postselected schemes~\cite{tichy2013entanglement,krenn2017entanglement,blasiak2019entangling,barros2020entangling,chin2021graph,lee2022entangling},  while viable with relatively simple circuits, come with a notable drawback. Since the creation of the target entangled state cannot be confirmed until the detection of photons, they render the target state a less suitable resource in quantum information processing. Ref.~\cite{adcock2018hard} also offers evidence that postselected entanglement gates are unable to cover the entire space of multipartite entanglements.

In contrast, heralded schemes~\cite{barz2010heralded,papp2009characterization,zeuner2018integrated,li2021heralded,le2021heralded} adopts a different strategy by incorporating ancillary photons and modes as ``heralds" to signal the successful generation of the anticipated target states. This allows for the discrimination of experimental runs that produce the desired target states without directly measuring those states. As a result, the entanglement generated through heralded schemes can be harnessed as a useful resource for quantum computations. However, the heralded schemes generally demand additional particles and modes for the resource of heralding, making them more intricate to design than postselected ones. 

To overcome this difficulty of heralded schemes, a graph approach to heralded entanglement generation is proposed in Ref.~\cite{chin2024shortcut}, providing a systematic strategy to design the schemes by boson subtractions. Ref.~\cite{chin2024shortcut} imposed correspondence relations between elements of bipartite graphs (bigraphs) and many-particle systems including boson  annihilation operators, which is an extension of the graph picture introduced in Ref.~\cite{chin2021graph}. To distinguish from other graphical approaches in quantum information and computation~\cite{van2020zx,biamonte2019lectures,hein2006entanglement,krenn2017entanglement}, we call our approach the graph picture of linear quantum networks, or \emph{linear quantum graph (LQG) picture}.

Based on the LQG picture,  Ref.~\cite{chin2024shortcut} provided various schemes to generate the multipartite entanglement of bosons with the sculpting protocol~\cite{karczewski2019sculpting,zaw2022sculpting}, which generates an $N$-partite entangled state by applying tailored $N+K$ single-boson subtraction operators (which is dubbed
the \emph{sculpting operator}) with $K$ ancillary bosons. In the LQG picture, all the sculpting operators are represented as bigraphs (which are dubbed \emph{sculpting} bigraphs). By leveraging the advantageous mathematical properties of the graphs, Ref.~\cite{chin2024shortcut} has found essential entanglement generation schemes including qubit $N$-partite GHZ and W states, $N=3$ Type 5 entangled states (the superposition of $N=3$ GHZ and W states~\cite{acin2000generalized}), and qudit $N$-partite GHZ states. 

On the other hand, we need one intermediate step to connect the sculpting bigraphs (hence sculpting operators) with the actual heralded circuits for generating multipartite entanglement: heralded operators for subtracting single bosons from a given multi-boson state. Once we know such operators (which we name ``subtractors'' here), \emph{we can automatically design heralded schemes for multipartite entangled states by assembling the operators according to the guidence of sculpting bigraphs.}

In this study, we demonstrate how to design heralded subtractors employing linear optical elements of polarized photons. Building upon this result, we establish translation rules for mapping sculpting bigraph elements to linear optical systems. By these rules, we can construct linear optical circuits that are the counterparts to any sculpting bigraphs for generating entanglement.
To showcase the practicality of our method, we use the sculpting bigraphs in Ref.~\cite{chin2024shortcut} to devise heralded schemes for the qubit $N$-partite GHZ state with $2N$ , $N$-partite W state, and $N=3$ Type 5 state. 

Our GHZ generating circuit can be applied to an arbitrary $N$-partite case with $2N$ photons, hence advantageous to schemes in Refs.~\cite{varnava2008good,gubarev2020improved}.   
The W generating scheme requires $2N+1$ single photons and linear operators, which is significantly more feasible than schemes in Refs.~\cite{ozdemir2011optical,li2020w} that use  $W$ states of small sizes as resources. The $N=3$ Type 5 state generating scheme is previously unreported  to our knowledge.
These are a part of schemes that our method can propose. The translation rules from sculpting bigraphs to linear optical schemes can be applied to any sculpting bigraph schemes for suitable entanglement generations. 

This work is organanized as follows: Section~\ref{review} explains the concept of sculpting protocol and its linear quantum graph (LQG) picture. Section~\ref{translation_rules} provides the translation rules from sculpting bigraphs to linear optical elements with the introduction of heralded subtraction operators (subtractors) in linear optics, by which we can construct heralded linear optical circuits that create GHZ, W, and the superposition of $N=3$ GHZ and W states in Section~\ref{circuits}. Section~\ref{discussions} summarized the physical implications of our research and further applications.  

\section{Sculpting protocol and its linear quantum graph (LQG) picture}~\label{review}

In this section, we review the concept of sculpting protocol~\cite{karczewski2019sculpting} and its directed bigraph representation~\cite{chin2024shortcut} in the graph picture of linear quantum networks (LQG picture). We call such bigraphs ``sculpting bigraphs''. Among those bigraphs, we present the definition of their most useful class: effective perfect matching (EPM) bigraphs, which directly correspond to sculpting operators that generate multipartite entanglement in our setup.

\subsection{Sculpting protocol}\label{sculpting protocol}
The sculpting protocol~\cite{karczewski2019sculpting,chin2024shortcut} exploits the indistinguishability of identical bosons and spatially overlapped subtraction operators (i.e., a single-boson subtraction operator is a superposition of subtractions on different spatial modes) to generate multipartite entanglement of bosons. 
	
In our physical setup, each boson in the $j$th spatial mode ($j\in \{1,2,\cdots, N\}$) has an internal degree of freedom of 2 dimension $s$ $(\in \{0,1\})$. Therefore boson creation  (annihilation) operators are denoted as $\ha_{j,s}^\dagger$ ($\ha_{j,s}$).
The sculpting protocol for generating multi-partite entanglement consists of three steps: 
\begin{enumerate}
    \item Initial state preparation: We prepare a product state of $2N$ bosons in $N$ spatial modes $|Sym_N\>$, of which each boson has different states (either spatial or internal) with each other: 
	\begin{align}\label{initial}
		|Sym_{N}\> &\equiv \ha^\dagger_{1,0}\ha^\dagger_{1,1} \ha^\dagger_{2,0}\ha^\dagger_{2,1} \cdots 
		\ha^\dagger_{N,0}\ha^\dagger_{N,1}|vac\>  \nn \\
		&= \prod_{j=1}^N(\ha^\dagger_{j,0}\ha^\dagger_{j,1})|vac\>. 
	\end{align} 
 
Aside from the state $|Sym_{N}\>$, some sculpting schemes require an ancillary system of $K$ $(\geq 0)$ spatial modes that has a single particle of the same internal state (here we choose $0$), 
 \begin{align}\label{Anc}
		|Anc_{K}\> &\equiv \ha^\dagger_{N+1,0} \ha^\dagger_{N+2,0} \cdots 
		\ha^\dagger_{N+K,0}|vac\>  \nn \\
		&= \prod_{j=1}^K(\ha^\dagger_{N+j,0})|vac\>. 
	\end{align} 
The whole initial state is $|Sym_{N}\>|Anc_{K}\>$ with the main (ancillary) system of $N$ ($K$) spatial modes, which becomes $|Sym_N\>$ when $K=0$.  
\item Operation: To the initial state $|Sym_{N}\>|Anc_{K}\>$,  we apply the \emph{sculpting operator}, which is written in the most general form as 	\begin{align}\label{annihilation}
		&	\prod_{l=1}^{N+K}\sum_{j=1}^{N+K}(k^{(l)}_{j ,0}\ha_{j,0} + k^{(l)}_{j,1}\ha_{j,1} ) \nn \\
		&\equiv \prod_{l=1}^{N+K}\hat{A}^{(l)} \equiv \hat{A}_{N+K}. \nn \\
		& \quad (k^{(l)}_{j,s} \in \mathbb{C}~\textrm{and}~  \sum_{j,s}|k^{(l)}_{j,s}|^2 =1).
	\end{align}
 where $\hat{A}^{(l)}$ is a single-boson subtraction operator. 
 The sculpting operator $\hat{A}_{N+K}$ must satisfy a crucial condition: it should eliminate exactly a single particle in each spatial mode. Following Ref.~\cite{chin2024shortcut}, we call it the \emph{no-bunching condition}. It ensures that the resulting overall state $|\P\>_{fin}$ is composed of states where each spatial mode in the main system contains one boson\footnote{Understanding the sculpting schemes with heralded schemes, the boson that remains in the mode contains the qubit information, while the subtracted boson serves as a heralding.}.  
 
 \item  Final state: We now obtain the final state by calculating 
	\begin{align}\label{final_state}
		|\P\>_{fin} = \hat{A}_{N+K}|Sym_{N}\>|Anc_{K}\> 
	\end{align}
 and verify whether it is genuinely multipartite entangled, i.e., the state cannot be separable under any bipartition of the given system~\cite{walter2016multipartite}.  
\end{enumerate}
$ $\\
The main challenge in determining $\hat{A}_N$ for a particular entangled state arises primarily from the no-bunching restriction.
As shown in Ref.~\cite{chin2024shortcut}, and to be briefed in the following subsection, the graph method serves as a potent tool to overcome this limitation.

\subsection{Directed bipartite graph representation of bosonic systems in LQG picture}

In Ref.~\cite{chin2024shortcut}, an undirected bigraph mapping was introduced (see Appendix~\ref{undirected_representation} for a brief summary of mapping), which was useful for finding several sculpting schemes in the work. One can consult Appendix A of Ref.~\cite{chin2021graph} for a simple glossary to the graph theory. While the undirected bigraph representation includes all the crucial information on the sculpting operators that can generate entanglement, we need a more comprehensive $directed$ bigraph representation to embrace the initial state and the sculpting operator in the LQG picture, which was first suggested in Appendix A of Ref.~\cite{chin2024shortcut} and will be more thoroughly explained here. We construct all our schemes based on the directed bigraph representation in this paper as it has a more direct and intuitive relation with heralded linear optical circuits. In the directed bigraph picture, the complete information on the sculpting protocol from the preparation to the final state is encoded in a graph, which we call a \emph{sculpting directed bigraph}. 


Table I enumerates the list of our correspondence relations from fundamental elements of many-boson systems to directed bipartite graphs. In the directed bigraph mapping, both creation and annihilation operators correspond to unlabelled vertices (dots), which are distinguished by the direction of edges attached to them.

\begin{widetext}
\begin{center}
		\begin{table}[t]
			\begin{tabular}{|l|l|}
				\hline
				\textbf{Many-particle systems}                & \textbf{Directed bipartite Graph $G_b = (U\cup V, E)$}        \\ \hline \hline 
					Spatial modes & Labelled vertices (circles, \encircle{$j$} $\in U$) \\ \hline					
					Creation operators & Unlabelled vertices (dots, $\bullet\in V$) with $outgoing$ edges.  \\ \hline
				
					Annihilation operators & Unlabelled vertices (dots, $\bullet\in V$) with $incoming$ edges \\ \hline	
					Spatial distributions of operators &  Directed edges $\in$ $E$ \\ \hline 				Probability amplitude $\a_j^{(l)} $ & Edge weight $\a_j^{(l)} $ \\ \hline 
Internal state $\p_j^{(l)} $ & Edge weight $\p_j^{(l)}$ (sometimes replaced with colors) \\ \hline 
				\end{tabular}
   \caption{Correspondence relations of many-particle systems to directed bipartite graphs }  
   \label{mapping}
			\end{table}
		\end{center} 
\end{widetext}

 The initial state $|Sym_N\>$ is now drawn  as	
 \begin{align}\label{initial_directed}
 |Sym_N\>= \prod_{j=1}^N(\ha^\dagger_{j,0}\ha^\dagger_{j,1})|vac\>= 
\begin{gathered}
\includegraphics[width=1.6cm]{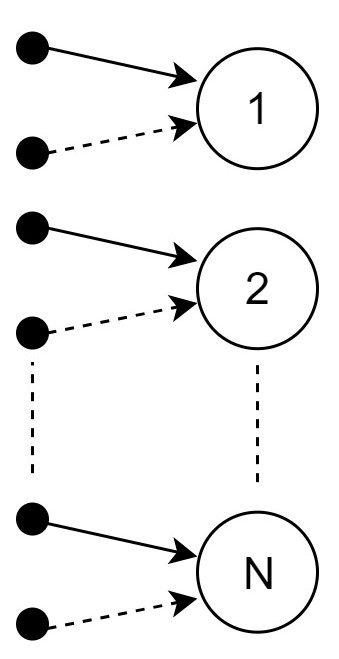}
\end{gathered}
	\end{align}		(the internal state edge weights $\{|0\>,|1\>,|+\>=\frac{1}{\sqrt2}(|0\>+|1\>),|-\>=\frac{1}{\sqrt2}(|0\>-|1\>) \}$ are denoted as edge colors $\{\textrm{Solid Black, Dotted Black, Red, Blue}\}$ respectively as in Ref.~\cite{chin2024shortcut}).

The directed bigraphs present a clear diagrammatic understanding of the following crucial identities
\begin{align}\label{qubit_identities}
\forall j & \in\{1,2,\cdots, N\}, \nn \\
		&\ha_{j,\pm}\ha^\dagger_{j,0}\ha^\dagger_{j,1}|vac\> = \pm\ha^\dagger_{j,\pm}|vac\>,  \nn \\
  &\ha_{j,+}\ha_{j,-}\ha^\dagger_{j,0}\ha^\dagger_{j,1}|vac\> = 0. 
\end{align} 

The first identity of Eq.~\eqref{qubit_identities} is expressed with directed bigraphs as
\begin{align}\label{qubit_identities_1}
\includegraphics[width=.4\textwidth]{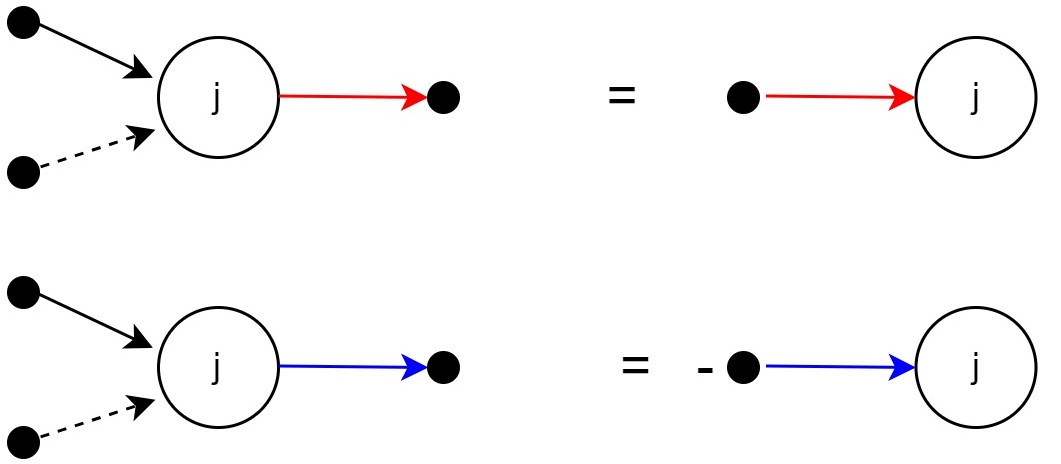}
\end{align}
and the second identity as
\begin{align}\label{qubit_identities_2}
\begin{gathered}
 \includegraphics[width=.18\textwidth]{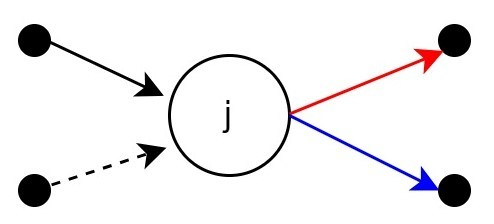}   
\end{gathered}
 =~0.
\end{align}
It is worth mentioning that  for any $n>1$
\begin{align}\label{qubit_identities_3}
    \ha_{j,0}^{n} \ha^\dagger_{j,0}|vac\> =\ha_{j,1}^{n} \ha^\dagger_{j,1}|vac\> =0
\end{align} corresponds to
\begin{align}\label{qubit_identities_4}
 \includegraphics[width=.47\textwidth]{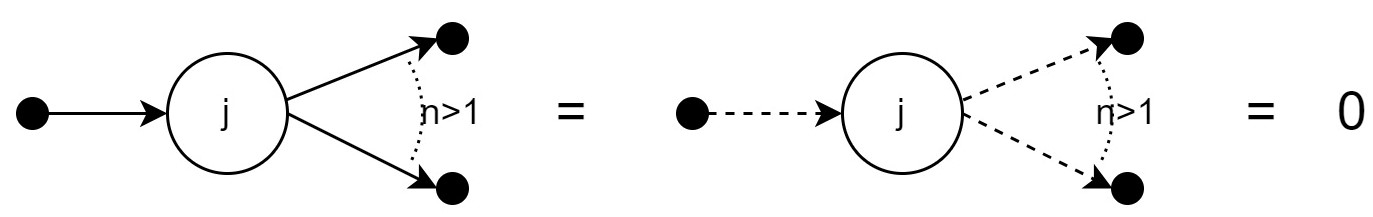}  
\end{align}
The above identities play an essential role in finding sculpting directed bigraphs for generating genuine entanglement by defining a special type of graphs, i.e., \emph{effective perfect matching (EPM) directed bigraphs}~\cite{chin2024shortcut}.    

\begin{figure*}[t] 		
			\includegraphics[width=15cm]{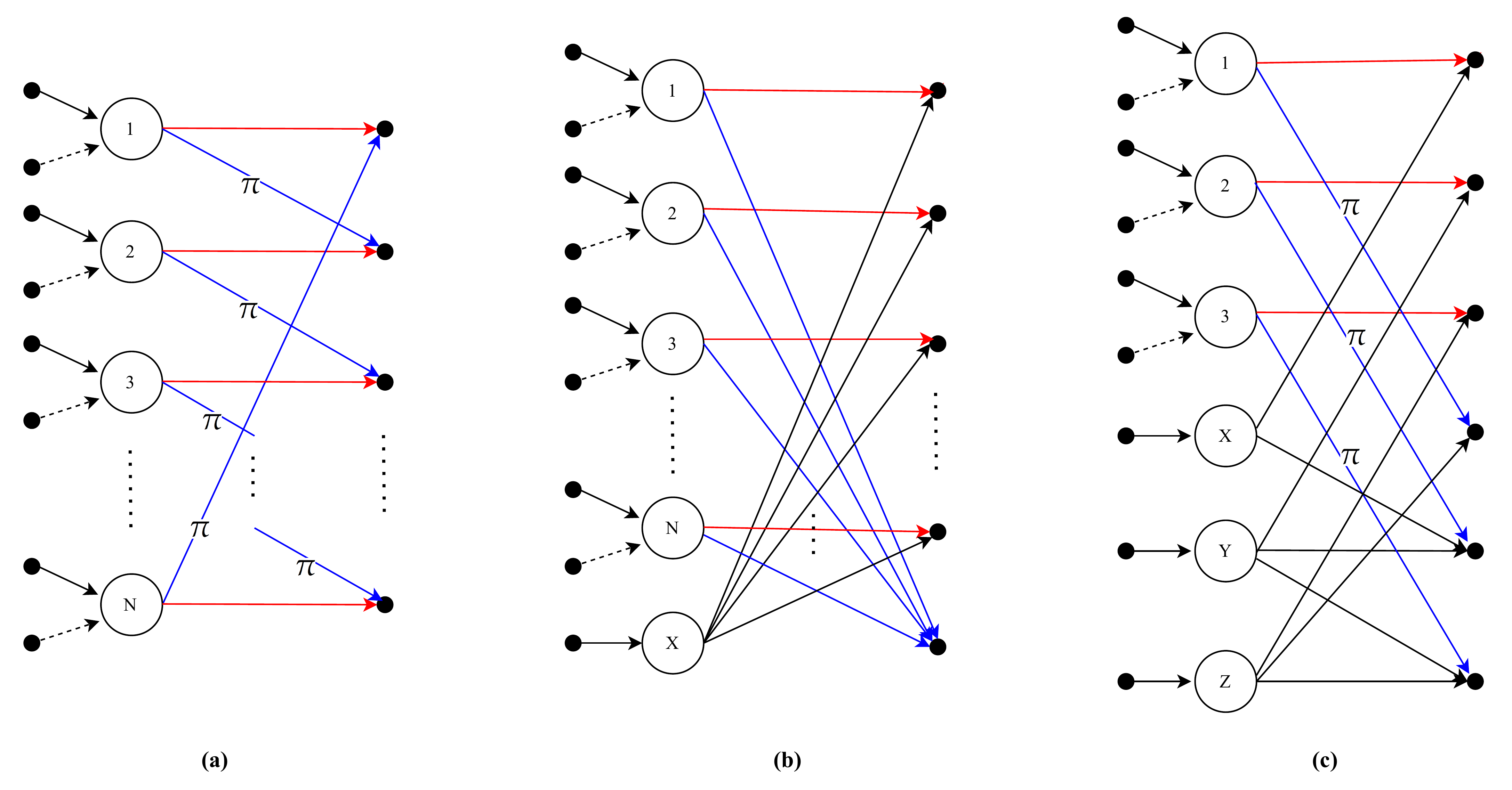}
			\caption{Three EPM directed bigraphs that corresponds to (a) GHZ-state, (b) W-state, and (c) $N=3$ Type 5 entangled states (a superposition of $N=3$ GHZ and W states, see Ref.~\cite{acin2000generalized} for a more rigorous definition) introduced in Ref.~\cite{chin2024shortcut}. Note that the probability amplitude weights are omitted for all of graphs except for the blue edges in (c) that have $\pi$. This is a \emph{simplified notation}, which can be applied to the cases when all the absolute values of probability amplitude weights are equal and normalized among those which go to the same dots. The edge weights are omitted when the phase is 0, and only phases are denoted otherwise.}
   \label{fig_sculpting_bigraphs}
\end{figure*}

\subsection{EPM directed bigraphs for generating entanglement}

We define EPM bigraphs as follows:

\begin{definition}
(EPM directed bigraphs) If all the edges of a bigraph attach to the circles as one of the following bigraphs
\begin{align}\label{EPM}
\begin{gathered}
    \includegraphics[width=.4\textwidth]{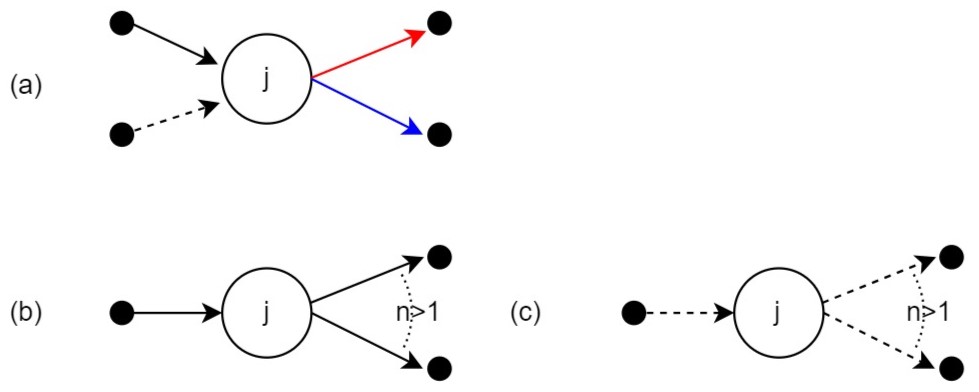}, 
\end{gathered}
\end{align} 
then it is an effective perfect matching (EPM) bigraph.
\end{definition}
A crucial property of an EPM bigraph is that the final state generated by the corresponding sculpting operator is fully determined by the perfect matchings (PMs) of the bigraph (a set of disjoint edges that cover all the vertices in a given bigraph, see Appendix A of Ref.~\cite{chin2021graph}), which we can show using identities~\eqref{qubit_identities_1} and \eqref{qubit_identities_2}~\cite{chin2024shortcut}. We can directly check that all the sculpting bigraphs for generating genuine qubit entanglement in Ref.~\cite{chin2024shortcut} (see \eqref{sculpting_bigraphs_und}) correspond to EPM  bigraphs.

\paragraph*{Remark 1.---} The main system and ancillary system consist of different type of bigraphs in Eq.~\eqref{EPM}.  Bigraph (a) of \eqref{EPM} corresponds to a spatial mode in the main system and (b) $or$ (c) to  a spatial mode in the ancillary system. Since the ancillary modes always has the same internal state, we need to choose between (b) and (c).
Following the definition of the ancillary system $|Anc_K\>$~\eqref{Anc}, we set that an EPM bigraph consists of circles of (a) and (b). See Fig.~\ref{fig_sculpting_bigraphs} (b) and (c) in which all the vertices are attached to the circles either as (a) or (b).


\paragraph*{Remark 2.---} Comparing the EPM directed bigraphs in Fig.~\ref{fig_sculpting_bigraphs} with the undirected EPM bigraphs in \eqref{sculpting_bigraphs_und}, we can easily see that the right hand side of the directed ones in Fig.~1 (a), (b), and (c)  without direction are identical to the three graphs in \eqref{sculpting_bigraphs_und} of Appendix~\ref{undirected_representation}. This is by the fact that the information on the sculpting operators is on the right hand side of the directed graphs, which is solely represented in the undirected graphs in \eqref{sculpting_bigraphs_und}.
This shows that the essential part in a EPM directed bigraph is the right hand side of the graph from the circles as it represents the sculpting operator that generates entanglement.

\paragraph*{Remark 3.---}
It is also worth mentioning the fundamental conceptual difference between the tensor network notation~\cite{orus2014practical,biamonte2019lectures} and our LQG picture. The tensor network notation is a graphical mapping of distinguishable qudits (states and effects) and operations (processes) among them that can correspond to some type of nonlocality. On the other hand, the physical elements that the LQG picture represents are discrete spatial modes and identical particles (bosons or fermions) that has some dynamical relations to them, as shown in Tables~\ref{mapping}.
 These elements cannot be considered qubit or qudit until they are encoded within the context of quantum information. For example, the initial state denoted as Eq.~\eqref{initial_directed} cannot be expressed as a multi-qubit state. We can state that the LQG picture describes crude physical elements that can be combined to play the role of qubits. 

\section{Translation rules from sculpting bigraph elements to linear optical elements}\label{translation_rules}

Ref.~\cite{chin2024shortcut} exploited the special property of EPM bigraphs (Property 2 of Section IV in Ref.~\cite{chin2024shortcut}) to find schemes for genuinely entangled states, which are enumerated in Fig.~\ref{fig_sculpting_bigraphs}. To design counterparts in linear optical circuits to the EPM bigraphs that generate the same entagled states, we need to know how each element of the graphs is translated to linear optical operators.
A prerequisite for the translation rules from graphs to circuits is to provide heralded linear optical operators (which we name \emph{subtractors}) that play the role of subtraction operators, because they are the fundamental components of EPM sculpting bigraphs. 

In this section, we explain how such subtractors in linear optical networks are obtained. Ref.~\cite{chin2024shortcut} presents a special set of subtractors that can only be applied to sculpting bigraphs whose edges are attached to the circles in the form of (a) in \eqref{EPM} 
(without ancilla), however it cannot be applied to general cases  including (b) or (c) in \eqref{EPM} that corresponds to ancillary systems. Here we provide a comprehensive set of translation rules from sculpting bigraphs to linear optical elements that can be applied to any type of EPM sculpting bigraphs.  

\subsection{EPM bigraph elements}

An arbitrary EPM directed bigraph that corresponds to a sculpting operator of a $(N+K)$-partite system ($N$ spatial modes in the main system and $K$ spatial modes in the ancillary system) consists of the following elements:\\
$ $\\
1.~$N$ circles (labelled vertices) to which edges are attached as
            \begin{align}\label{main}
   \begin{gathered}
\includegraphics[width=.15\textwidth]{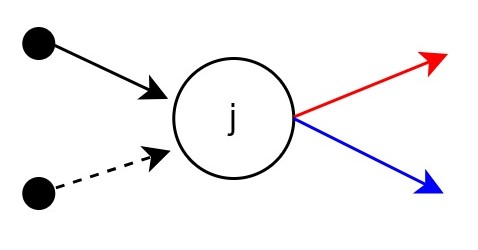}   
   \end{gathered} 
   \end{align} \\
2.~$K$ circles to which edges are attached as 
 \begin{align}\label{ancillae}
   \begin{gathered}
\includegraphics[width=.15\textwidth]{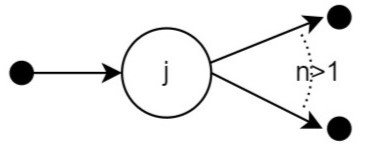}        
   \end{gathered} 
    \end{align}  \\
3.~Dots (unlabelled vertices) with $q$ incoming edges from the $N$ circles of the form~\eqref{main} and $k$ incoming edges from the $K$ circles of the form~\eqref{ancillae}:
      \begin{align} 
   \begin{gathered}
\includegraphics[width=.08\textwidth]{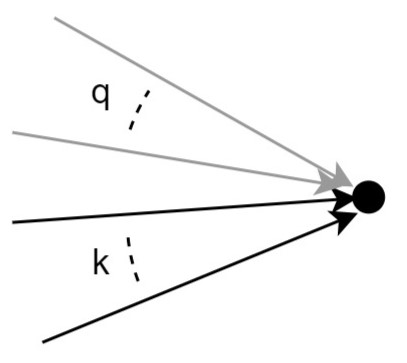} 
   \end{gathered} 
    \end{align} 
where the gray  edges denote that they can have any color between red and blue.


We can check that $K=0$ for the GHZ graph, $K=1$ for the W graph, and $K=3$ for the $N=3$ Type-5 graph. 

\subsection{Subtraction operators in linear optics with heralding}\label{subtraction_operators}

To design associated optical circuits from the EPM sculpting bigraphs, we need to define heralded operators for subtracting single bosons, which we name \emph{subtractors}. 
In this section, we will show that such operators can be constructed in linear optics with compositions of following operators: 
$ $\\
1. Polarizing beam splitters (PBSs) that transform photons as
\begin{align}
\includegraphics[width=.5\textwidth]{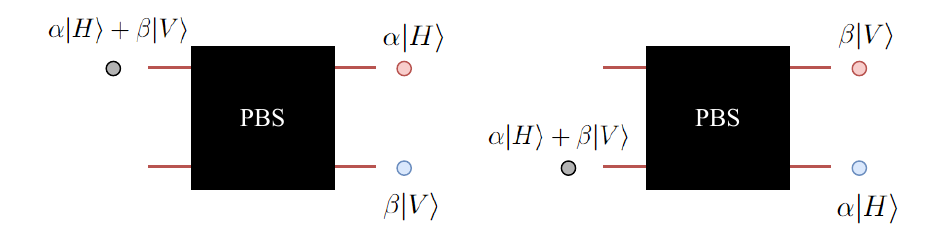} \nn 
\end{align}\\
2. Half-wave plates with
HWP: $\{H,V\} \leftrightarrow \{D,A\}$ and \underline{HWP}: $\{H,V\} \leftrightarrow \{A,D\}$ 

3. $n$-partite multiports that Fourier-transform the spatial mode states of photons (a $2$-partite port corresponds to a beam splitter (BS))


We start from the heralded schemes for the identities~\eqref{qubit_identities_1}, whose operational expressions become
\begin{align}
    \ha_{\pm}\ha^\dagger_0\ha^\dagger_1|vac\> = \pm \ha^\dagger_{\pm}|vac\>
\end{align} as this is a fundamental element for EPM directed bigraphs. 
 
We first consider the heralded optical circuit for taking $\ha_{+}$, which can be rewritten as
\begin{align}
\ha_{+}\frac{(\ha^{\dagger 2}_+ - \ha^{\dagger 2}_-)}{2}|vac\>  = \ha^\dagger_{+}|vac\>.  
\end{align} This operation is implemented with a linear optical scheme as  
\begin{align}\label{subtraction+}
    \includegraphics[width=8cm]{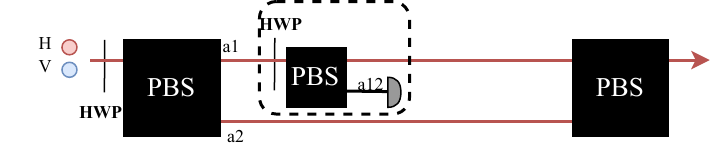}
\end{align} by encoding the internal degree of freedom with the photon polarization as
\begin{align}
\{|0\>, |1\>\} \cong \{|D\>, |A\> \},~~~
\{|+\>, |-\>\} \cong \{|H\>, |V\> \}. 
\end{align}
\textbf{Step-by-step explanation}
\begin{enumerate}
    \item  The 1st HWP: Preparation of the initial state
    \begin{align}
        \ha^\dagger_{H}\ha^\dagger_V \to \ha^\dagger_{D}\ha^\dagger_A =  \frac{1}{2}(\ha^{\dagger 2}_H -\ha^{\dagger 2}_V)
    \end{align}
    \item The 1st PBS: Division of the particles with different internal states into two distinguishable spatial modes: 
    \begin{align}
            \frac{1}{2}(\ha^{\dagger 2}_H - \ha^{\dagger 2}_V) \to \frac{1}{2}(\ha^{\dagger 2}_{1,H} - \ha^{\dagger 2}_{2,V}) 
    \end{align}
where 1 and 2 denote the upper and lower paths of PBS respectively.
    \item The dashed box: Subtraction of $\ha^\dagger_{1,H}$ by a measurement in which exactly one photon is registered in mode 12
    \begin{align}
        \frac{1}{2}(\ha^{\dagger 2}_{1,H} - \ha^{\dagger 2}_{2,V})  \to \ha^\dagger_{1,H}.
    \end{align} Here a photon number resolving detector is required to execute the postselection of the one-photon state.
    We can see that the measurement of one photon at the detector 12 corresponds to the subtraction of one photon. 
    \item The last PBS: Mergence of the particle paths by which the photon always arrives at one spatial mode: \begin{align}
        \ha^\dagger_{1 H} \to \ha^\dagger_{H}
    \end{align}
\end{enumerate}
Note that the dashed box corresponds to a particle subtraction by postselection, which we name a \emph{subtractor}. 

In the above process, we can obtain the same result by  simply attenuating the lower wire instead of merging two paths of wires by the last PBS, which however is needed for the general cases when the heralded subtraction is overlapped among several spatial modes.

The circuit for taking $\ha_{-}$ can be designed similarly as
\begin{align}\label{subtraction-}
    \includegraphics[width=8cm]{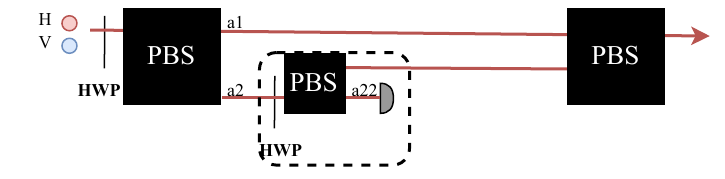}
\end{align}

We can generalize the above subtractor 
to spatially overlapped subtractions (i.e., a single-boson subtraction operator $\hat{A}^{(l)}$ is a superposition of subtractions on different spatial modes)  from $k$ different spatial modes by superposing the paths of detected particles.
Hence, for example, a subtraction $(\ha_{1+} -\ha_{2-})$ from two modes $1$ and $2$ corresponds to
\begin{align}\label{subtraction_indet}
\includegraphics[width=5cm]{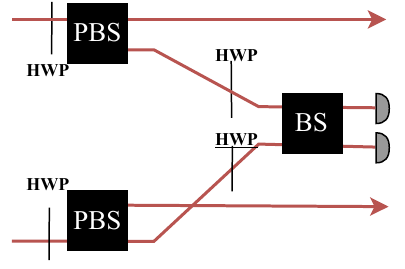},
\end{align} 
where the incoming wires are attached to $11$ and  $22$ by the internal states of the annihilation operators. 
The beam splitter (denoted as BS) is always balanced in our work and hence considered the 2-partite port.  
In the above subtractor, the two open wires go back to the spatial modes from which those wires came and be merged by PBSs  as in~\eqref{subtraction+} and \eqref{subtraction-}.  

In the most general sense, we can subtract a photon from more than $l$ ($\geq 2$) modes, some from the main system (two photons per mode) and some from the ancillary system (one photon per mode). For the case, \emph{we place a PBS on the path from the main system and make a measurement in the mutually unbiased basis using $K$-partite port}. Based on this principle, we can build the subtractors needed for the EPM sculpting bigraphs in our work:
\begin{align}
    \includegraphics[width=8cm]{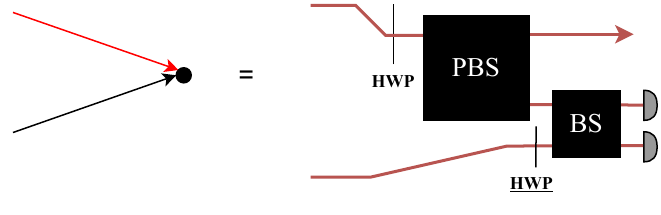} 
\end{align} and 
\begin{align}\label{W_N3}
    \includegraphics[width=8cm]{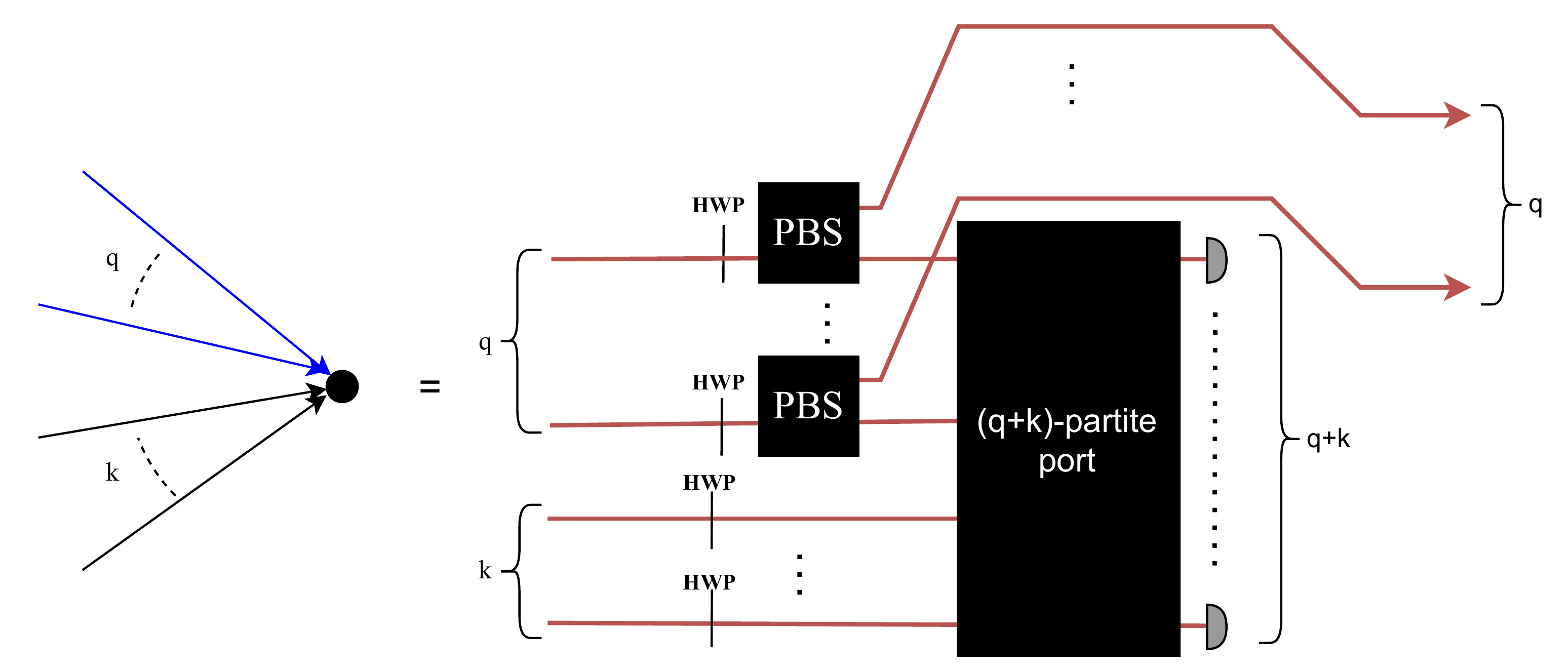}     
\end{align} 
where we postselect the case when only one photon arrives at the detectors of the above two subtractors. In the subtractor~\eqref{W_N3}, the $q$ open wires go back to the spatial modes from which those wires came and be merged by PBSs  as in~\eqref{subtraction+} and \eqref{subtraction-}.

On the other hand, we can also consider an alternative circuit that plays the roles of circuits~\eqref{subtraction+} and~\eqref{subtraction-} simultaneously as
\begin{align}
    \includegraphics[width=8cm]{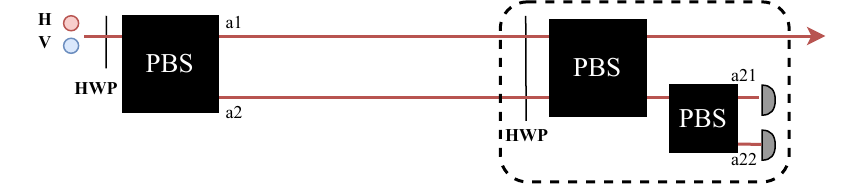},
\end{align}  which is proposed and analyzed in Ref.~\cite{chin2024shortcut}.
We can use this alternative subtraction operator when the one-boson subtraction operator is a superposition of two annihilation operators and their internal state is orthogonal to each other. $(\ha_+ -\hb_-)$ is an example of such an operator. Then, 
\begin{align}
    \includegraphics[width=6cm]{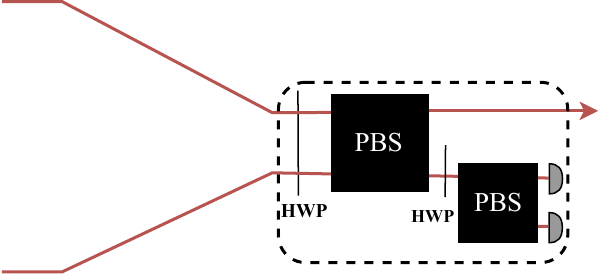}
\end{align} subtracts one photon in a spatially overlapped way when a photon arrives at either uppper or lower detector.
In the above operator, we need an HWP between two PBSs, with which we can generate the spatially overlapped subtractor of different internal states from difference input modes. 
We define the above operator as an \emph{optimized subtractor} as it needs less PBSs than the subtractor~\eqref{subtraction_indet}, denoting it as
\begin{align}
\includegraphics[width=8cm]{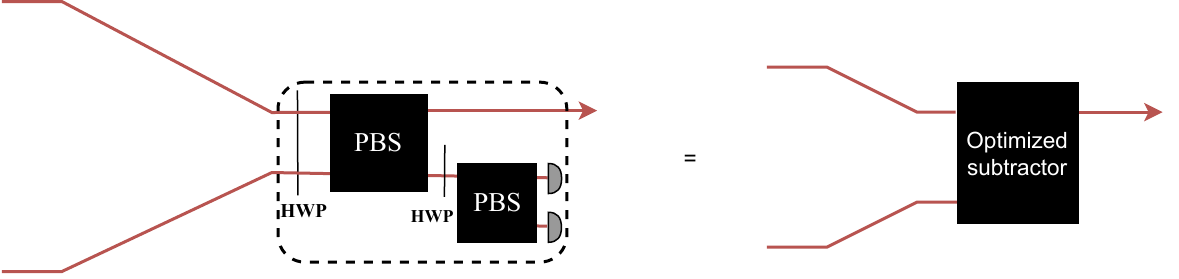}.
\end{align}
In the graph picture, we obtain the following equality:
\begin{align}
\includegraphics[width=8cm]{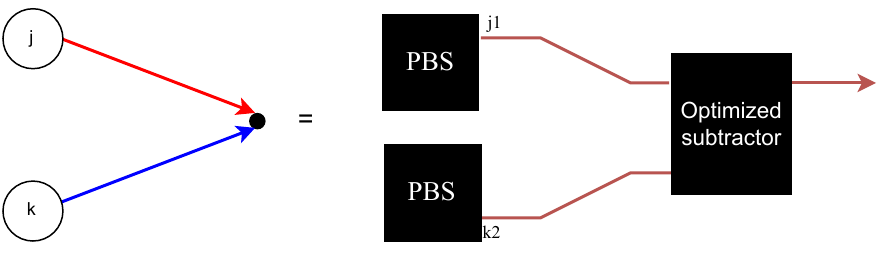}
\end{align}
It is worth noting that the optimized subtractor works in a difference way from the Type II fusion measurement~\cite{browne2005resource} in spite of their structural similarity. While only one particle comes into each arm of Type II gate, the optimized subtractor is installed in our setup to subtract a single photon from the bunching state of two photons.

\begin{widetext}
\begin{figure*}[t] 		
			\includegraphics[width=14cm]{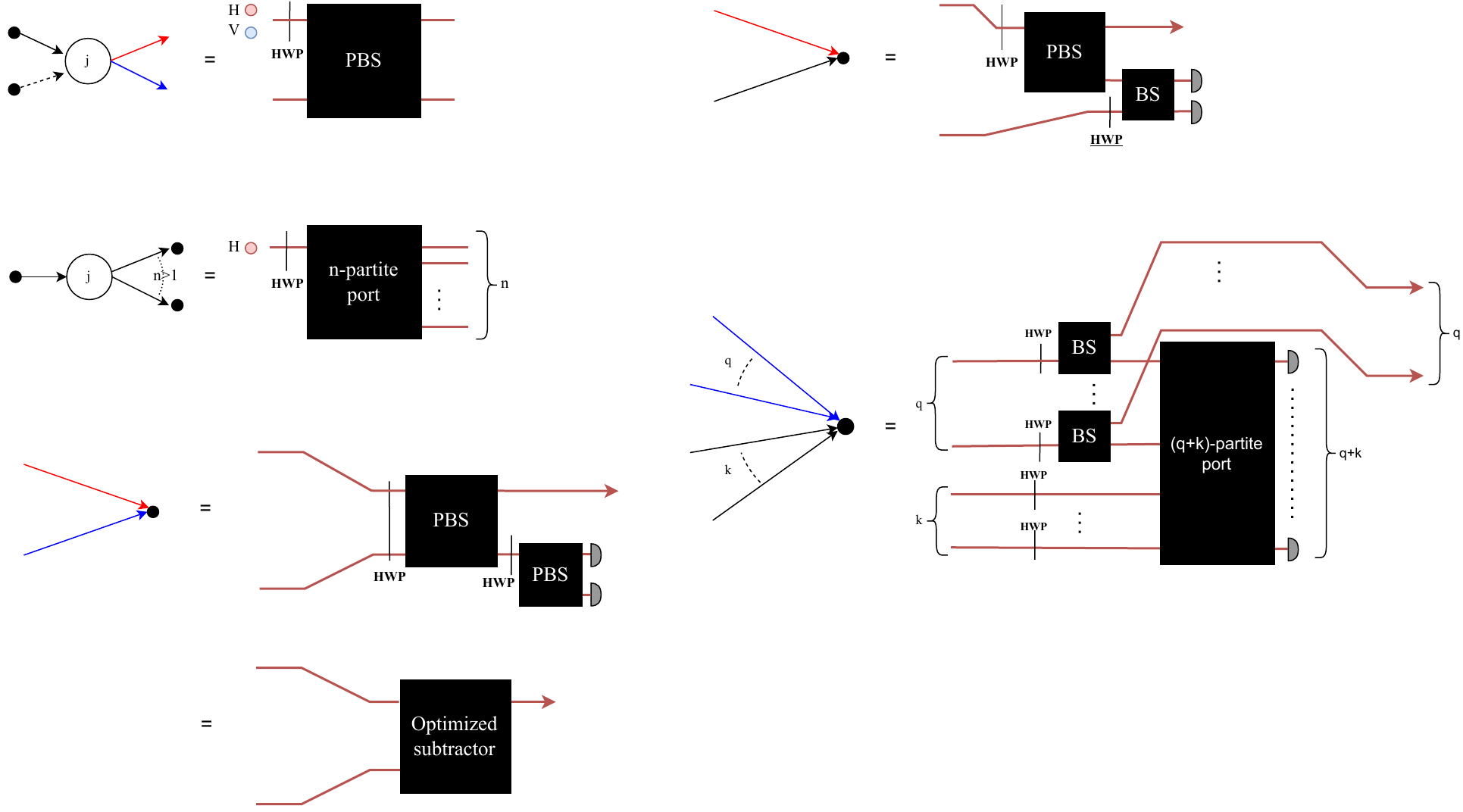}
			\caption{Translation rules from EPM bigraph elements to linear optical elements. In the  above rules, if the edge is red (blue) from the $j$th circle, the corresponding wire is attached to the upper (lower) mode of the $j$th PBS. And in the bottom right rule, the $q$ open wires go back to the spatial modes from which those wires came and be merged by PBSs.}
   \label{fig_translation_rules}
\end{figure*}
\end{widetext}

With the subtractors as building blocks, we can enumerate all the elements that construct EPM directed bigraphs and present translation rules of them to linear optical elements. 
All the translation rules from EPM sculpting bigraph elements to linear optical elements that we have explained are enumerated in Fig.~\ref{fig_translation_rules}.  
Therefore, an EPM bigraph that generates a multipartite entangled state serves as a blueprint for designing a linear optical circuit that generates the same entangled state, as will be clarified in the next section.





\section{Linear optical circuits}\label{circuits}

Using the translation rules in Fig.~\ref{fig_translation_rules}, we can directly design a linear optical circuit that corresponds to any EPM bigraph. In this section, we showcase how the EPM sculpting bigraphs for GHZ, W, and the superposition of $N=3$ GHZ and W states in Fig.~\ref{fig_sculpting_bigraphs} are translated to linear optical circuits for the same target states.  

\subsection{$N$-partite GHZ state (Fig.~\ref{fig_sculpting_bigraphs} (a))} 


\begin{figure}
    \centering
     \includegraphics[width=.3\textwidth]{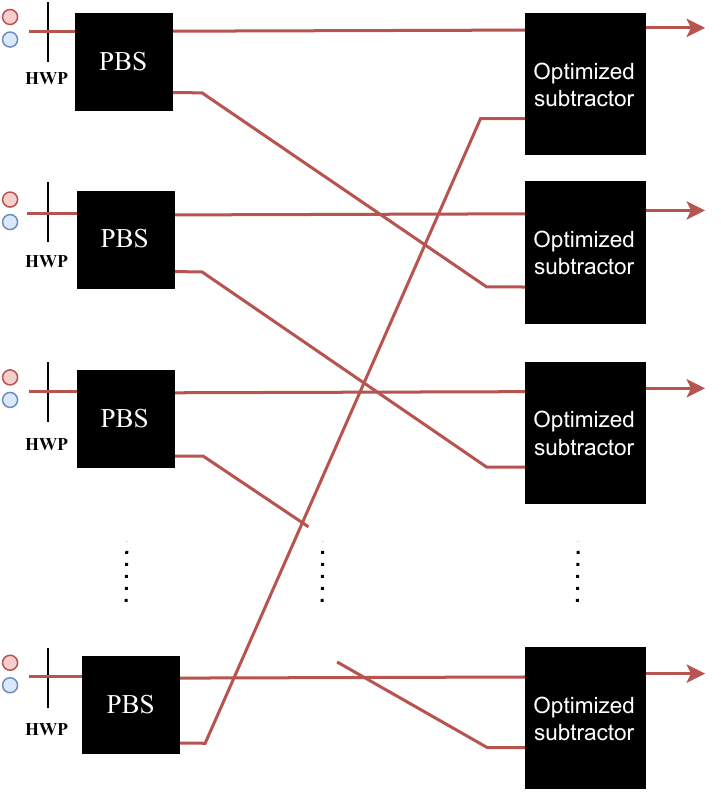} 
    \caption{$N$-partite GHZ state generation scheme with $2N$ photons and the success probability $P_{suc}= \frac{1}{2^{2N-1}}$.}
    \label{fig_GHZ}
\end{figure}

Following the translation rules, we simply obtain a linear optical circuit for the GHZ state as Fig.~\ref{fig_GHZ}.
Comparing the circuit in Fig.~\ref{fig_GHZ} with the GHZ sculpting bigraph (Fig.~\ref{fig_sculpting_bigraphs} (a)), we can see that the photon paths are connected following the structure of the bigraph.
The explicit calculation for showing that the above circuit generates the GHZ state is given in Appendix~\ref{GHZ_cal}. The success probability $P_{suc}$ is  $\frac{1}{2^{2N-1}}$ with feed-forward\footnote{The feed-forward scheme involves measurement result dependent single qubit operations, which are not challenging quantum operations in themselves. However, the scheme requires fast data processing to apply these operations in real-time based on measurement outcomes~\cite{luiz2021fiber}.
}.

We can compare our scheme with other heralded GHZ-state-generating  schemes~\cite{varnava2008good,gubarev2020improved,zou2005scheme} using various criteria. Before we get into the discussion, it is worth noting that all the schemes mentioned here require the same number of single photons for the same number of parties ($2N$ photons for the $N$-partite GHZ state).
 First, our scheme can be directly generalized to any $N$-partite case unlike Ref.~\cite{gubarev2020improved} that suggests schemes for $N=2$ and 3 based on numerical optimizations or Ref.~\cite{varnava2008good} for the $N=3$ GHZ entanglement generation using fusion gates (see Refs.~\cite{maring2024versatile,cao2024photonic} for the experimental results). Second, contrary to schemes in Refs.~\cite{gubarev2020improved,zou2005scheme} that depend on the dual-rail encoding structure, our scheme can be installed with any kind of qubit encoding (polarization, dual-rail, etc.) based on the sculpting bigraph. See  Appendix~\ref{polarization_to_dual} for the transformation from  polarization encoding circuits to dual-rail encoding circuits.
Finally, we can also compare the optical component complexity\footnote{It is worth noting that a direct comparison of the complexity or number of elements  among different schemes is subtle in this case. The biggest reason is that each scheme requires a different type of devices, e.g., polarizing beam splitters (PBS) or beam splitters (BS) of various transmissivities. In addition, Refs. [23,24] is just about the N=3 GHZ state scheme, while ours presents a scheme for an arbitrary N.}. Our scheme requires $3N$ PBSs and $2N$ detectors, which is simpler than the scheme in Ref.~\cite{zou2005scheme} that requires $4N$ BSs and $2N$ detectors. Additionally comparing with the schemes in Refs.~\cite{varnava2008good,gubarev2020improved} for $N=3$ case, Ref.~\cite{varnava2008good} requires the simplest circuit elements (5 PBSs and 6 detectors) and Ref.~\cite{gubarev2020improved} the most complicated (12 BSs of various transitivities and 6 detectors).



\subsection{$N$-partite W state (Fig.~\ref{fig_sculpting_bigraphs}, (b))}

\begin{figure}
    \centering
     \includegraphics[width=.5\textwidth]{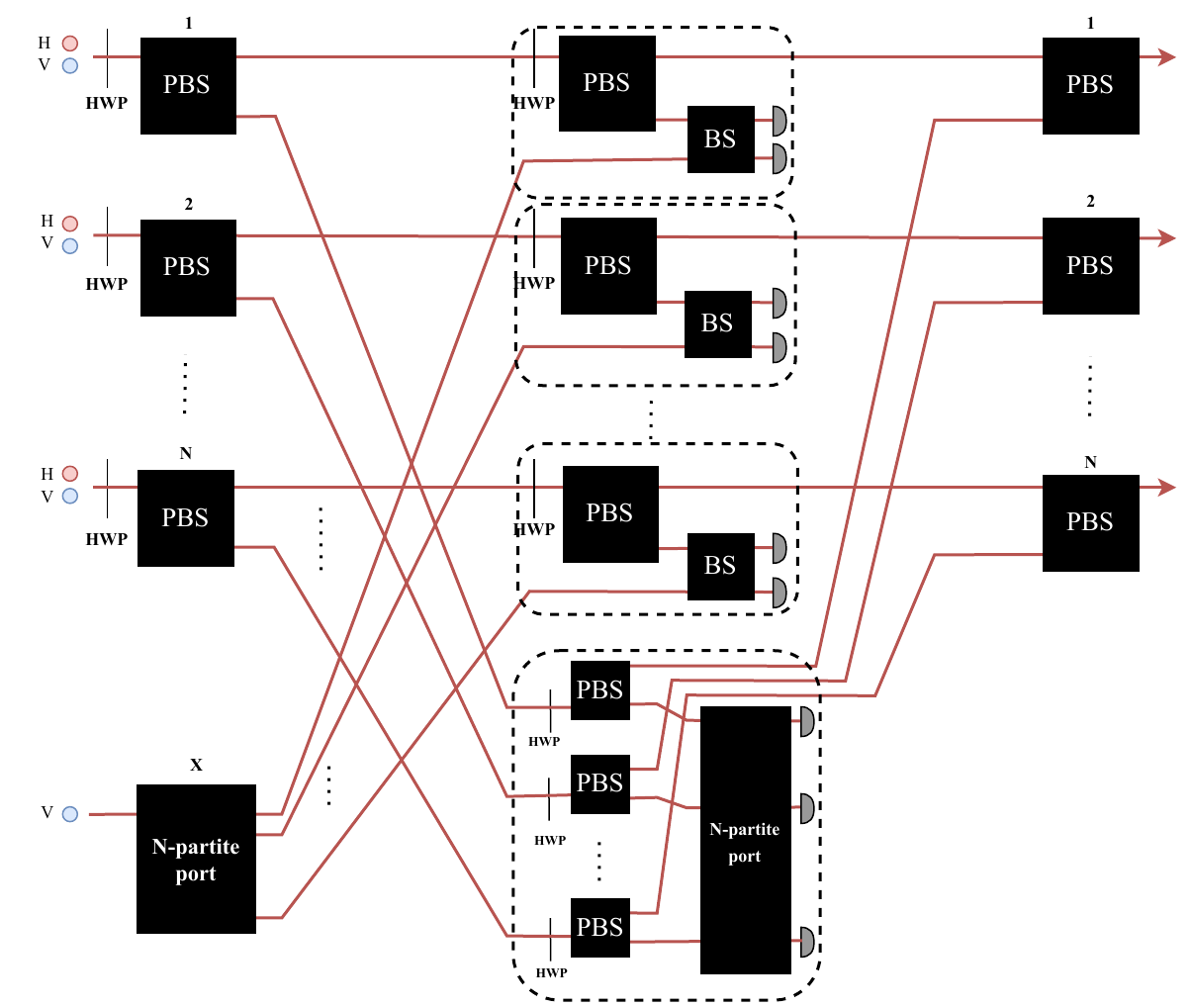} 
    \caption{$N$-partite W state generation scheme with $2N+1$ photons and the success probability $P_{suc}= \frac{1}{2^{2N}}$.}
    \label{fig_W}
\end{figure}
A linear optical scheme for the W generating EPM bigraph (Fig.~\ref{fig_sculpting_bigraphs}, (b)) from the translation rules is as in Fig.~\ref{fig_W}. 
Considering each dashed box (spatially overlapped subtractors) as a dot, we can see that the wires are connected following the structure of the W sculpting bigraph.
Appendix~\ref{W_cal} provides an explicit calculation for showing that the above circuit generates the W state with $P_{suc}=\frac{1}{2^{2N}}$ with feed-forward.

To our knowledge, no linear optical heralded W-state scheme for arbitrary $N$ party with single-photon sources has been proposed. Since our scheme requires $2N+1$ single-photons, it is much more feasible than those schemes which exploit fusion gates and small size of W states as resources for generating $N$-partite W states~\cite{ozdemir2011optical,li2020w}. 
\subsection{$N=3$ Type 5 state (Fig.~\ref{fig_sculpting_bigraphs} (c))}
\begin{figure}
    \centering
     \includegraphics[width=.5\textwidth]{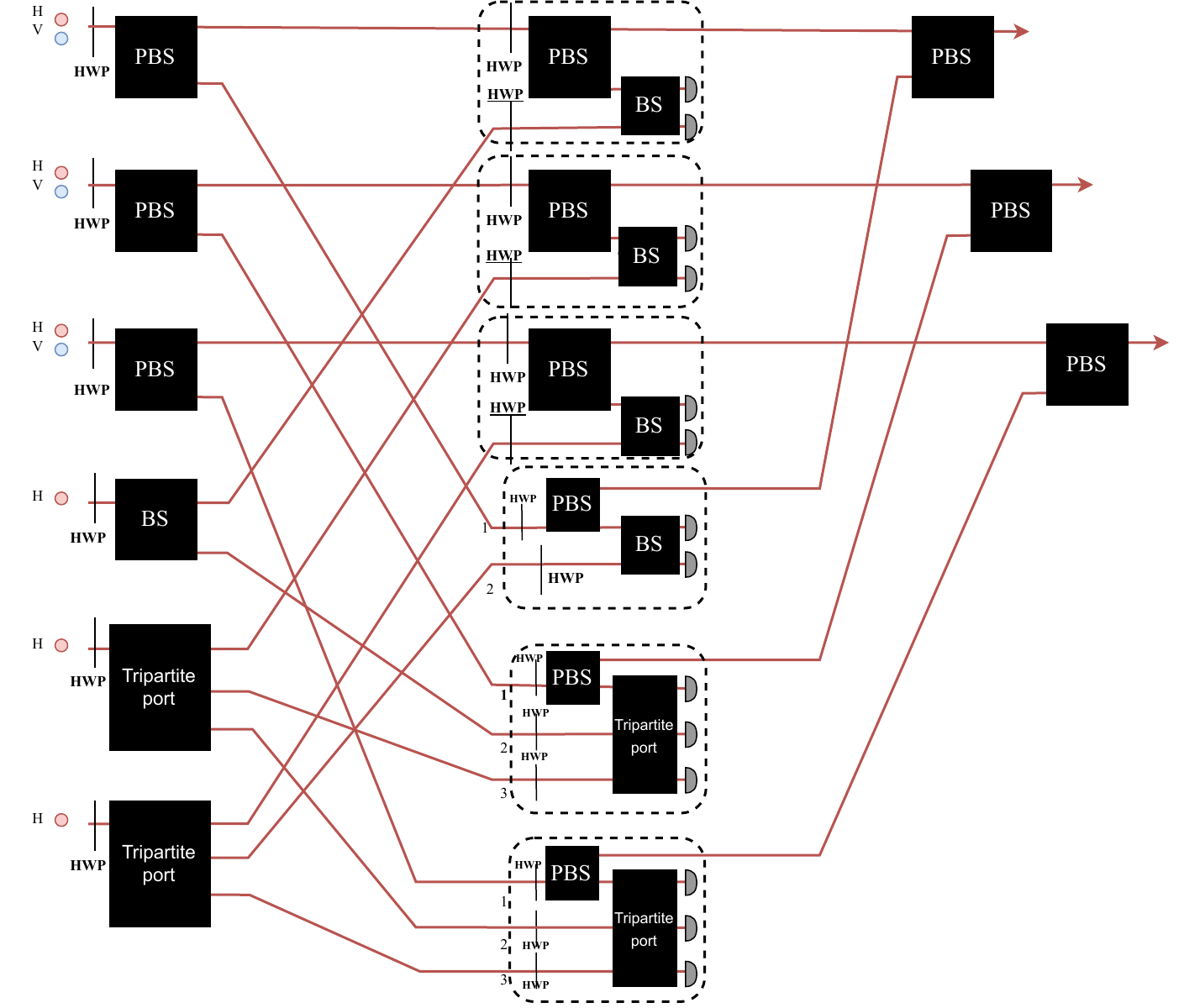} 
    \caption{$N=3$ type 5 state generation scheme with 9 photons and the success probability $P_{suc}= \frac{5}{3^22^7}$.}
    \label{fig_N=3}
\end{figure}

The GHZ and W states stand out as representative states of two distinct categories of $N=3$ genuine entanglement. Therefore, a superposed state of GHZ and W states can be a useful resource that cannot be generated by local operations and classical communications (LOCCs) to either GHZ or W states. We can express such states as a superposition of five local bases product states ($N=3$ Type 5 
 state~\cite{acin2000generalized})
\begin{align}
 \{|+++\>, |-++\>,|--+\>, |+-+\>,|---\>\}. 
\end{align} We can obtain such a state with an EPM bigraph Fig.~\ref{fig_sculpting_bigraphs} (c), whose corresponding optical circuit is presented in Fig.~\ref{fig_N=3}.
Considering each dashed box as a dot, we can see that the wires are connected following the structure of the $N=3$ Type 5 entangled state sculpting bigraph. See Appendix~\ref{N3_cal} for 
the explicit calculation. $P_{suc}=5/(3^2 2^7)$ with feed-forward. To our knowledge, no linear optical heralded scheme for the $N=3$ Type 5 state has been proposed before.

\section{Discussions}\label{discussions}

By establishing a set of translation rules from sculpting bigraphs into linear optical circuits, we have proposed a systematic method to design heralded circuits for the multipartite entanglment of bosons by combining the results in Ref.~\cite{chin2024shortcut} with translation rules in Fig.~\ref{fig_translation_rules}.
Although our work primarily addresses optical circuits employing polarization encoding, it is worth noting that \emph{conceptually equivalent circuits in dual-rail encoding} can be directly designed~(Appendix~\ref{polarization_to_dual}).

Our GHZ scheme provides a general framework for an arbitrary number of parties and is compatible with any form of qubit encoding. It also does not have any trusted third-party for heralding the target state, which is advantageous in quantum communication. See 
 Ref.~\cite{zo2024heralded} for a more detailed explanation on this aspect of our GHZ scheme. Ref.~\cite{zo2024heralded} also provides an analysis on the effect of channel losses on our scheme. For the cases of W and $N=3$ Type 5 state, no linear optical heralded scheme has been proposed to our knowledge.  Given the latest experimental demonstrations~\cite{maring2024versatile,cao2024photonic} of generating the $N=3$ GHZ state based on the scheme in Ref.~\cite{varnava2008good}, we believe that our proposed scheme is also realizable in the near future. Alternatively, one can also use online quantum computing platforms such as Quandela~\cite{maring2024versatile} that offer the linear operations involving 6 ideal photons with the integrated photonics by
adopting our circuits into dual-rail encoding circuits. 

The methodology discussed herein is applicable to any entanglement generating sculping protocols using EPM sculpting bigraphs, e.g., caterpillar graph state generating EPM bigraphs proposed in Ref.~\cite{chin2023linear}. This implies that \emph{our translation rules substantially simplify the task of finding heralded schemes for generating entanglement into that of finding EPM bigraphs that correspond to suitable sculpting operators. }

We can extend our discussion to qudit entanglements such as qudit $N$-partite GHZ states~\cite{chin2024exponentially} that we can also generate with the sculpting protocol as given in Ref.~\cite{chin2024shortcut}.  We can choose another photonic degree of freedom as the internal state for encoding qudits, e.g., the orbital angular momentum (OAMs). For the case, the PBSs are replaced with OAM beam splitters~\cite{zou2005scheme} and HWPs with OAM-only Fourier transformation operators~\cite{kysela2020fourier}.


$ $\\
\section*{Acknowledgements}
SC is grateful to Prof. William J. Munro, Prof. Ana Belen Sainz, and Prof. Jung-Hoon Chun for their support on this research. This research is funded by
National Research Foundation of Korea (NRF, RS-2023-00245747, 2021M3H3A103657313, 2022M3K4A1094774, and 2023M3K5A1094805),  Korea Institute of Science and Technology (2E31021), and Foundation for Polish Science (IRAP project, ICTQT, contract no.2018/MAB/5, co-financed by EU within Smart Growth Operational Programme)

\bibliographystyle{unsrtnat}
\bibliography{mybib}

\begin{thebibliography}{37}
\providecommand{\natexlab}[1]{#1}
\providecommand{\url}[1]{\texttt{#1}}
\expandafter\ifx\csname urlstyle\endcsname\relax
  \providecommand{\doi}[1]{doi: #1}\else
  \providecommand{\doi}{doi: \begingroup \urlstyle{rm}\Url}\fi

\bibitem[Plenio and Virmani(2007)]{plenio2007introduction}
Martin~B Plenio and Shashank Virmani.
\newblock An introduction to entanglement measures.
\newblock \emph{Quantum Inf. Comput.}, 7\penalty0 (1):\penalty0 1--51, 2007.
\newblock URL \url{https://doi.org/10.26421/QIC7.1-2-1}.

\bibitem[Preskill(2012)]{preskill2012quantum}
John Preskill.
\newblock Quantum computing and the entanglement frontier.
\newblock \emph{arXiv preprint arXiv:1203.5813}, 2012.
\newblock URL \url{https://doi.org/10.48550/arXiv.1203.5813}.

\bibitem[Gisin et~al.(2002)Gisin, Ribordy, Tittel, and
  Zbinden]{gisin2002quantum}
Nicolas Gisin, Gr{\'e}goire Ribordy, Wolfgang Tittel, and Hugo Zbinden.
\newblock Quantum cryptography.
\newblock \emph{Reviews of modern physics}, 74\penalty0 (1):\penalty0 145,
  2002.
\newblock URL \url{https://doi.org/10.1103/RevModPhys.74.145}.

\bibitem[Tichy et~al.(2013)Tichy, de~Melo, Ku{\'s}, Mintert, and
  Buchleitner]{tichy2013entanglement}
Malte~C Tichy, Fernando de~Melo, Marek Ku{\'s}, Florian Mintert, and Andreas
  Buchleitner.
\newblock Entanglement of identical particles and the detection process.
\newblock \emph{Fortschritte der Physik}, 61\penalty0 (2-3):\penalty0 225--237,
  2013.
\newblock URL \url{https://doi.org/10.1002/prop.201200079}.

\bibitem[Krenn et~al.(2017)Krenn, Hochrainer, Lahiri, and
  Zeilinger]{krenn2017entanglement}
Mario Krenn, Armin Hochrainer, Mayukh Lahiri, and Anton Zeilinger.
\newblock Entanglement by path identity.
\newblock \emph{Physical Review Letters}, 118\penalty0 (8):\penalty0 080401,
  2017.
\newblock URL \url{https://doi.org/10.1103/PhysRevLett.118.080401}.

\bibitem[Blasiak and Markiewicz(2019)]{blasiak2019entangling}
Pawel Blasiak and Marcin Markiewicz.
\newblock Entangling three qubits without ever touching.
\newblock \emph{Scientific Reports}, 9\penalty0 (1):\penalty0 20131, 2019.
\newblock URL \url{https://doi.org/10.1038/s41598-019-55137-3}.

\bibitem[Barros et~al.(2020)Barros, Chin, Pramanik, Lim, Cho, Huh, and
  Kim]{barros2020entangling}
Mariana~R Barros, Seungbeom Chin, Tanumoy Pramanik, Hyang-Tag Lim, Young-Wook
  Cho, Joonsuk Huh, and Yong-Su Kim.
\newblock Entangling bosons through particle indistinguishability and spatial
  overlap.
\newblock \emph{Optics Express}, 28\penalty0 (25):\penalty0 38083--38092, 2020.
\newblock URL \url{https://doi.org/10.1364/OE.410361}.

\bibitem[Chin et~al.(2021)Chin, Kim, and Lee]{chin2021graph}
Seungbeom Chin, Yong-Su Kim, and Sangmin Lee.
\newblock Graph picture of linear quantum networks and entanglement.
\newblock \emph{Quantum}, 5:\penalty0 611, 2021.
\newblock URL \url{https://doi.org/10.22331/q-2021-12-23-611}.

\bibitem[Lee et~al.(2022)Lee, Pramanik, Hong, Cho, Lim, Chin, and
  Kim]{lee2022entangling}
Donghwa Lee, Tanumoy Pramanik, Seongjin Hong, Young-Wook Cho, Hyang-Tag Lim,
  Seungbeom Chin, and Yong-Su Kim.
\newblock Entangling three identical particles via spatial overlap.
\newblock \emph{Optics Express}, 30\penalty0 (17):\penalty0 30525--30535, 2022.
\newblock URL \url{https://doi.org/10.1364/OE.460866}.

\bibitem[Adcock et~al.(2018)Adcock, Morley-Short, Silverstone, and
  Thompson]{adcock2018hard}
Jeremy~C Adcock, Sam Morley-Short, Joshua~W Silverstone, and Mark~G Thompson.
\newblock Hard limits on the postselectability of optical graph states.
\newblock \emph{Quantum Science and Technology}, 4\penalty0 (1):\penalty0
  015010, 2018.
\newblock URL \url{https://doi.org/10.1088/2058-9565/aae950}.

\bibitem[Barz et~al.(2010)Barz, Cronenberg, Zeilinger, and
  Walther]{barz2010heralded}
Stefanie Barz, Gunther Cronenberg, Anton Zeilinger, and Philip Walther.
\newblock Heralded generation of entangled photon pairs.
\newblock \emph{Nature Photonics}, 4\penalty0 (8):\penalty0 553--556, 2010.
\newblock URL \url{https://doi.org/10.1038/nphoton.2010.156}.

\bibitem[Papp et~al.(2009)Papp, Choi, Deng, Lougovski, Van~Enk, and
  Kimble]{papp2009characterization}
Scott~B Papp, Kyung~Soo Choi, Hui Deng, Pavel Lougovski, SJ~Van~Enk, and
  HJ~Kimble.
\newblock Characterization of multipartite entanglement for one photon shared
  among four optical modes.
\newblock \emph{Science}, 324\penalty0 (5928):\penalty0 764--768, 2009.
\newblock URL \url{https://doi.org/10.1126/science.1172260}.

\bibitem[Zeuner et~al.(2018)Zeuner, Sharma, Tillmann, Heilmann, Gr{\"a}fe,
  Moqanaki, Szameit, and Walther]{zeuner2018integrated}
Jonas Zeuner, Aditya~N Sharma, Max Tillmann, Ren{\'e} Heilmann, Markus
  Gr{\"a}fe, Amir Moqanaki, Alexander Szameit, and Philip Walther.
\newblock Integrated-optics heralded controlled-not gate for
  polarization-encoded qubits.
\newblock \emph{npj Quantum Information}, 4\penalty0 (1):\penalty0 13, 2018.
\newblock URL \url{https://doi.org/10.1038/s41534-018-0068-0}.

\bibitem[Li et~al.(2021)Li, Gu, Qin, Wu, You, Wang, Schneider, H{\"o}fling,
  Huo, Lu, et~al.]{li2021heralded}
Jin-Peng Li, Xuemei Gu, Jian Qin, Dian Wu, Xiang You, Hui Wang, Christian
  Schneider, Sven H{\"o}fling, Yong-Heng Huo, Chao-Yang Lu, et~al.
\newblock Heralded nondestructive quantum entangling gate with single-photon
  sources.
\newblock \emph{Physical Review Letters}, 126\penalty0 (14):\penalty0 140501,
  2021.
\newblock URL \url{https://doi.org/10.1103/PhysRevLett.126.140501}.

\bibitem[Le et~al.(2021)Le, Asavanant, and An]{le2021heralded}
Dat~Thanh Le, Warit Asavanant, and Nguyen~Ba An.
\newblock Heralded preparation of polarization entanglement via quantum
  scissors.
\newblock \emph{Physical Review A}, 104\penalty0 (1):\penalty0 012612, 2021.
\newblock URL \url{https://doi.org/10.1103/PhysRevA.104.012612}.

\bibitem[Chin et~al.(2024{\natexlab{a}})Chin, Kim, and
  Karczewski]{chin2024shortcut}
Seungbeom Chin, Yong-Su Kim, and Marcin Karczewski.
\newblock Shortcut to multipartite entanglement generation: A graph approach to
  boson subtractions.
\newblock \emph{npj Quantum Information}, 10\penalty0 (1):\penalty0 67,
  2024{\natexlab{a}}.
\newblock URL \url{https://doi.org/10.1038/s41534-024-00845-6}.

\bibitem[van~de Wetering(2020)]{van2020zx}
John van~de Wetering.
\newblock {ZX}-calculus for the working quantum computer scientist.
\newblock \emph{arXiv preprint arXiv:2012.13966}, 2020.
\newblock URL \url{https://doi.org/10.48550/arXiv.2012.13966}.

\bibitem[Biamonte(2019)]{biamonte2019lectures}
Jacob Biamonte.
\newblock Lectures on quantum tensor networks.
\newblock \emph{arXiv preprint arXiv:1912.10049}, 2019.
\newblock URL \url{https://doi.org/10.48550/arXiv.1912.10049}.

\bibitem[Hein et~al.(2006)Hein, D{\"u}r, Eisert, Raussendorf, Nest, and
  Briegel]{hein2006entanglement}
Marc Hein, Wolfgang D{\"u}r, Jens Eisert, Robert Raussendorf, M~Nest, and H-J
  Briegel.
\newblock Entanglement in graph states and its applications.
\newblock \emph{arXiv preprint quant-ph/0602096}, 2006.
\newblock URL \url{https://doi.org/10.48550/arXiv.quant-ph/0602096}.

\bibitem[Karczewski et~al.(2019)Karczewski, Lee, Ryu, Lasmar, Kaszlikowski, and
  Kurzy{\'n}ski]{karczewski2019sculpting}
Marcin Karczewski, Su-Yong Lee, Junghee Ryu, Zakarya Lasmar, Dagomir
  Kaszlikowski, and Pawe{\l} Kurzy{\'n}ski.
\newblock Sculpting out quantum correlations with bosonic subtraction.
\newblock \emph{Physical Review A}, 100\penalty0 (3):\penalty0 033828, 2019.
\newblock URL \url{https://doi.org/10.1103/PhysRevA.100.033828}.

\bibitem[Zaw et~al.(2022)Zaw, Lasmar, Nguyen, Tseng, Matsukevich, Kaszlikowski,
  and Scarani]{zaw2022sculpting}
Lin~Htoo Zaw, Zakarya Lasmar, Chi-Huan Nguyen, Ko-Wei Tseng, Dzmitry
  Matsukevich, Dagomir Kaszlikowski, and Valerio Scarani.
\newblock Sculpting bosonic states with arithmetic subtractions.
\newblock \emph{New Journal of Physics}, 24\penalty0 (8):\penalty0 083023,
  2022.
\newblock URL \url{https://doi.org/10.1088/1367-2630/ac8305}.

\bibitem[Ac{\'\i}n et~al.(2000)Ac{\'\i}n, Andrianov, Costa, Jan{\'e}, Latorre,
  and Tarrach]{acin2000generalized}
Antonio Ac{\'\i}n, A~Andrianov, L~Costa, E~Jan{\'e}, JI~Latorre, and Rolf
  Tarrach.
\newblock Generalized schmidt decomposition and classification of
  three-quantum-bit states.
\newblock \emph{Physical Review Letters}, 85\penalty0 (7):\penalty0 1560, 2000.
\newblock URL \url{https://doi.org/10.1103/PhysRevLett.85.1560}.

\bibitem[Varnava et~al.(2008)Varnava, Browne, and Rudolph]{varnava2008good}
Michael Varnava, Daniel~E Browne, and Terry Rudolph.
\newblock How good must single photon sources and detectors be for efficient
  linear optical quantum computation?
\newblock \emph{Physical review letters}, 100\penalty0 (6):\penalty0 060502,
  2008.
\newblock URL \url{https://doi.org/10.1103/PhysRevLett.100.060502}.

\bibitem[Gubarev et~al.(2020)Gubarev, Dyakonov, Saygin, Struchalin, Straupe,
  and Kulik]{gubarev2020improved}
FV~Gubarev, IV~Dyakonov, M~Yu Saygin, GI~Struchalin, SS~Straupe, and SP~Kulik.
\newblock Improved heralded schemes to generate entangled states from single
  photons.
\newblock \emph{Physical Review A}, 102\penalty0 (1):\penalty0 012604, 2020.
\newblock URL \url{https://doi.org/10.1103/PhysRevA.102.012604}.

\bibitem[{\"O}zdemir et~al.(2011){\"O}zdemir, Matsunaga, Tashima, Yamamoto,
  Koashi, and Imoto]{ozdemir2011optical}
{\c{S}}~K {\"O}zdemir, Eiji Matsunaga, Toshiyuki Tashima, Takashi Yamamoto,
  Masato Koashi, and Nobuyuki Imoto.
\newblock An optical fusion gate for {W}-states.
\newblock \emph{New Journal of Physics}, 13\penalty0 (10):\penalty0 103003,
  2011.
\newblock URL \url{https://doi.org/10.1088/1367-2630/13/10/103003}.

\bibitem[Li et~al.(2020)Li, Zheng, Xu, Mao, and Wang]{li2020w}
Ke~Li, Dongliang Zheng, Wangqiong Xu, Huibing Mao, and Jiqing Wang.
\newblock {W} states fusion via polarization-dependent beam splitter.
\newblock \emph{Quantum Information Processing}, 19\penalty0 (11):\penalty0
  412, 2020.
\newblock URL \url{https://doi.org/10.1007/s11128-020-02898-w}.

\bibitem[Walter et~al.(2016)Walter, Gross, and Eisert]{walter2016multipartite}
Michael Walter, David Gross, and Jens Eisert.
\newblock Multipartite entanglement.
\newblock \emph{Quantum Information: From Foundations to Quantum Technology
  Applications}, pages 293--330, 2016.
\newblock URL \url{https://doi.org/10.1002/9783527805785.ch14}.

\bibitem[Or{\'u}s(2014)]{orus2014practical}
Rom{\'a}n Or{\'u}s.
\newblock A practical introduction to tensor networks: Matrix product states
  and projected entangled pair states.
\newblock \emph{Annals of physics}, 349:\penalty0 117--158, 2014.
\newblock URL \url{https://doi.org/10.1016/j.aop.2014.06.013}.

\bibitem[Browne and Rudolph(2005)]{browne2005resource}
Daniel~E Browne and Terry Rudolph.
\newblock Resource-efficient linear optical quantum computation.
\newblock \emph{Physical Review Letters}, 95\penalty0 (1):\penalty0 010501,
  2005.
\newblock URL \url{https://doi.org/10.1103/PhysRevLett.95.010501}.

\bibitem[Luiz~Zanin et~al.(2021)Luiz~Zanin, Jacquet, Spagnolo, Schiansky,
  Calafell, Rozema, and Walther]{luiz2021fiber}
Guilherme Luiz~Zanin, Maxime~J Jacquet, Michele Spagnolo, Peter Schiansky,
  Irati~Alonso Calafell, Lee~A Rozema, and Philip Walther.
\newblock Fiber-compatible photonic feed-forward with 99\% fidelity.
\newblock \emph{Optics Express}, 29\penalty0 (3):\penalty0 3425--3437, 2021.
\newblock URL \url{https://doi.org/10.1364/OE.409867}.

\bibitem[Zou and Mathis(2005)]{zou2005scheme}
XuBo Zou and W~Mathis.
\newblock Scheme for optical implementation of orbital angular momentum beam
  splitter of a light beam and its application in quantum information
  processing.
\newblock \emph{Physical Review A}, 71\penalty0 (4):\penalty0 042324, 2005.
\newblock URL \url{https://doi.org/10.1103/PhysRevA.71.042324}.

\bibitem[Maring et~al.(2024)Maring, Fyrillas, Pont, Ivanov, Stepanov, Margaria,
  Hease, Pishchagin, Lema{\^\i}tre, Sagnes, et~al.]{maring2024versatile}
Nicolas Maring, Andreas Fyrillas, Mathias Pont, Edouard Ivanov, Petr Stepanov,
  Nico Margaria, William Hease, Anton Pishchagin, Aristide Lema{\^\i}tre,
  Isabelle Sagnes, et~al.
\newblock A versatile single-photon-based quantum computing platform.
\newblock \emph{Nature Photonics}, pages 1--7, 2024.
\newblock URL \url{https://doi.org/10.1038/s41566-024-01403-4}.

\bibitem[Cao et~al.(2024)Cao, Hansen, Giorgino, Carosini, Zah{\'a}lka, Zilk,
  Loredo, and Walther]{cao2024photonic}
H~Cao, LM~Hansen, F~Giorgino, L~Carosini, P~Zah{\'a}lka, F~Zilk, JC~Loredo, and
  P~Walther.
\newblock Photonic source of heralded greenberger-horne-zeilinger states.
\newblock \emph{Physical Review Letters}, 132\penalty0 (13):\penalty0 130604,
  2024.
\newblock URL \url{https://doi.org/10.1103/PhysRevLett.132.130604}.

\bibitem[Zo et~al.(2024)Zo, Chin, and Kim]{zo2024heralded}
Wan Zo, Seungbeom Chin, and Yong-Su Kim.
\newblock Heralded optical entanglement distribution via lossy quantum
  channels: A comparative study.
\newblock \emph{arXiv preprint arXiv:2409.16622}, 2024.
\newblock URL \url{https://doi.org/10.48550/arXiv.2409.16622}.

\bibitem[Chin(2023)]{chin2023linear}
Seungbeom Chin.
\newblock From linear quantum system graphs to qubit graphs: Heralded
  generation of graph states.
\newblock \emph{arXiv preprint arXiv:2306.15148}, 2023.
\newblock URL \url{https://doi.org/10.48550/arXiv.2306.15148}.

\bibitem[Chin et~al.(2024{\natexlab{b}})Chin, Ryu, and
  Kim]{chin2024exponentially}
Seungbeom Chin, Junghee Ryu, and Yong-Su Kim.
\newblock Exponentially enhanced scheme for the heralded qudit
  greenberger-horne-zeilinger state in linear optics.
\newblock \emph{Phys. Rev. Lett.}, 133:\penalty0 253601, 2024{\natexlab{b}}.
\newblock URL \url{https://doi.org/10.1103/PhysRevLett.133.253601}.

\bibitem[Kysela et~al.(2020)Kysela, Gao, and Daki{\'c}]{kysela2020fourier}
Jaroslav Kysela, Xiaoqin Gao, and Borivoje Daki{\'c}.
\newblock Fourier transform of the orbital angular momentum of a single photon.
\newblock \emph{Physical Review Applied}, 14\penalty0 (3):\penalty0 034036,
  2020.
\newblock URL \url{https://doi.org/10.1103/PhysRevApplied.14.034036}.

\end{thebibliography}

\newpage
\onecolumngrid	
\appendix 

\section{Undirected bipartite graph representation of bosonic systems in LQG picture}\label{undirected_representation}

As we can see from Section~\ref{sculpting protocol}, essential information on generating multipartite entanglement with sculpting protocol lies in the sculpting operators. Therefore, we can always consider a simpler graph representation that only reflects the property of sculpting operators without the direction of edges, i.e., undirected bipartite representation~\cite{chin2024shortcut}. Here we briefly summarize the undirected representation of LQG picture and how they are directly related to the directed representation. 

	\begin{center} 
		\begin{table}[h]
			\begin{tabular}{|l|l|l|}
				\hline
				\textbf{Boson systems with sculpting operator} & \textbf{Bipartite Graph $G_b =(U\cup V, E)$}  \\
				\hline\hline 
				Spatial modes & Labelled vertices $\in$ $U$  \\ \hline 
				$\hat{A}^{(l)}$  ($l \in \{1,2,\cdots, N\}$) & Unlabelled vertices $\in$  $V$ \\ \hline 
				Spatial distributions of $\hat{A}^{(l)}$  & Edges  $\in$ $E$ \\ \hline 
				Probability amplitude $\a_j^{(l)} $ & Edge weight $\a_j^{(l)} $ \\ \hline 
				Internal state $\p_j^{(l)} $ & Edge weight $\p_j^{(l)}$  \\
				\hline 
			\end{tabular}
			\label{dict}
			\caption{Correspondence relations from a sculpting operator to a undirected bipartite graph}
   \label{mapping_undirected}
		\end{table} 
	\end{center} 
Then the identities 
	\begin{align}		&\ha_{j,+}\ha_{j,-}\ha^\dagger_{j,0}\ha^\dagger_{j,1}|vac\> = 0, \nn \\
		& \ha_{j,0}^{n} \ha^\dagger_{j,0}\ha^\dagger_{j,1}|vac\> =  \ha_{j,1}^{n} \ha^\dagger_{j,0}\ha^\dagger_{j,1}|vac\> =0. ~~~(n\geq 2)		
	\end{align}
 are translated into the undirected  bigraph language as
 \begin{align}
 \begin{gathered}
     \includegraphics[width=6cm]{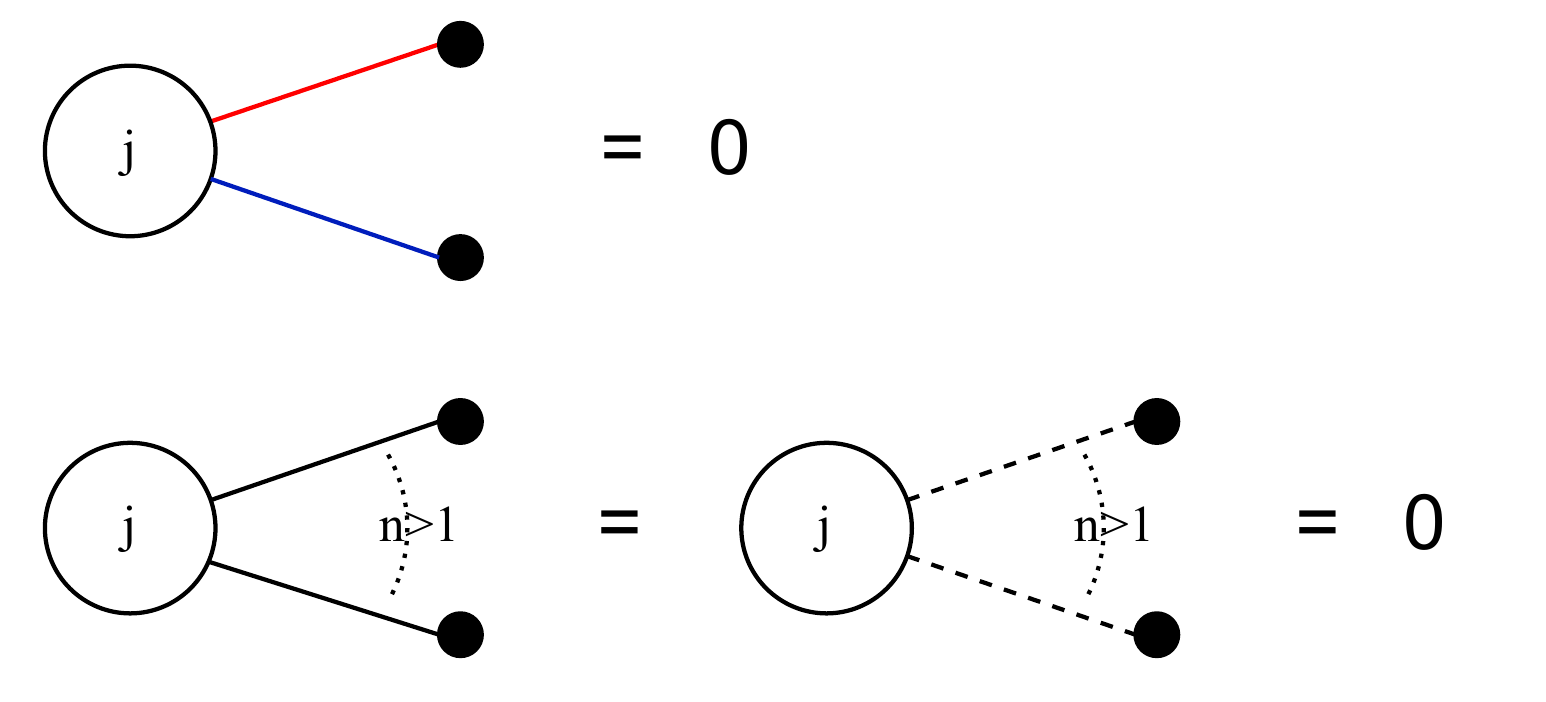}   
 \end{gathered}.
 \end{align}
Note that the above graphical relations correspond to \eqref{qubit_identities_2} and \eqref{qubit_identities_4} of the directed bigraph representation.

Undirected EPM bigraphs are now defined as follows.  If all the edges of a bigraph attach to the circles as one of the following undirected bigraphs
 \begin{align}
 \begin{gathered}
     \includegraphics[width=8cm]{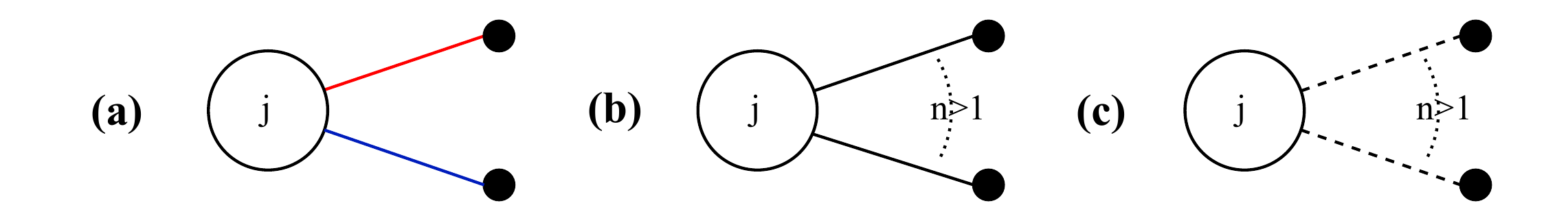}    
 \end{gathered},
 \end{align} then it is an undirected EPM bigraph. 

Three undirected EPM biraphs that corresponds to (a) GHZ state, (b) W-state, (c) $N=3$ Type 5 entangled state are proposed in Ref.~\cite{chin2024shortcut} as follows: 
\begin{align}\label{sculpting_bigraphs_und}
\begin{gathered}
     \includegraphics[width=14cm]{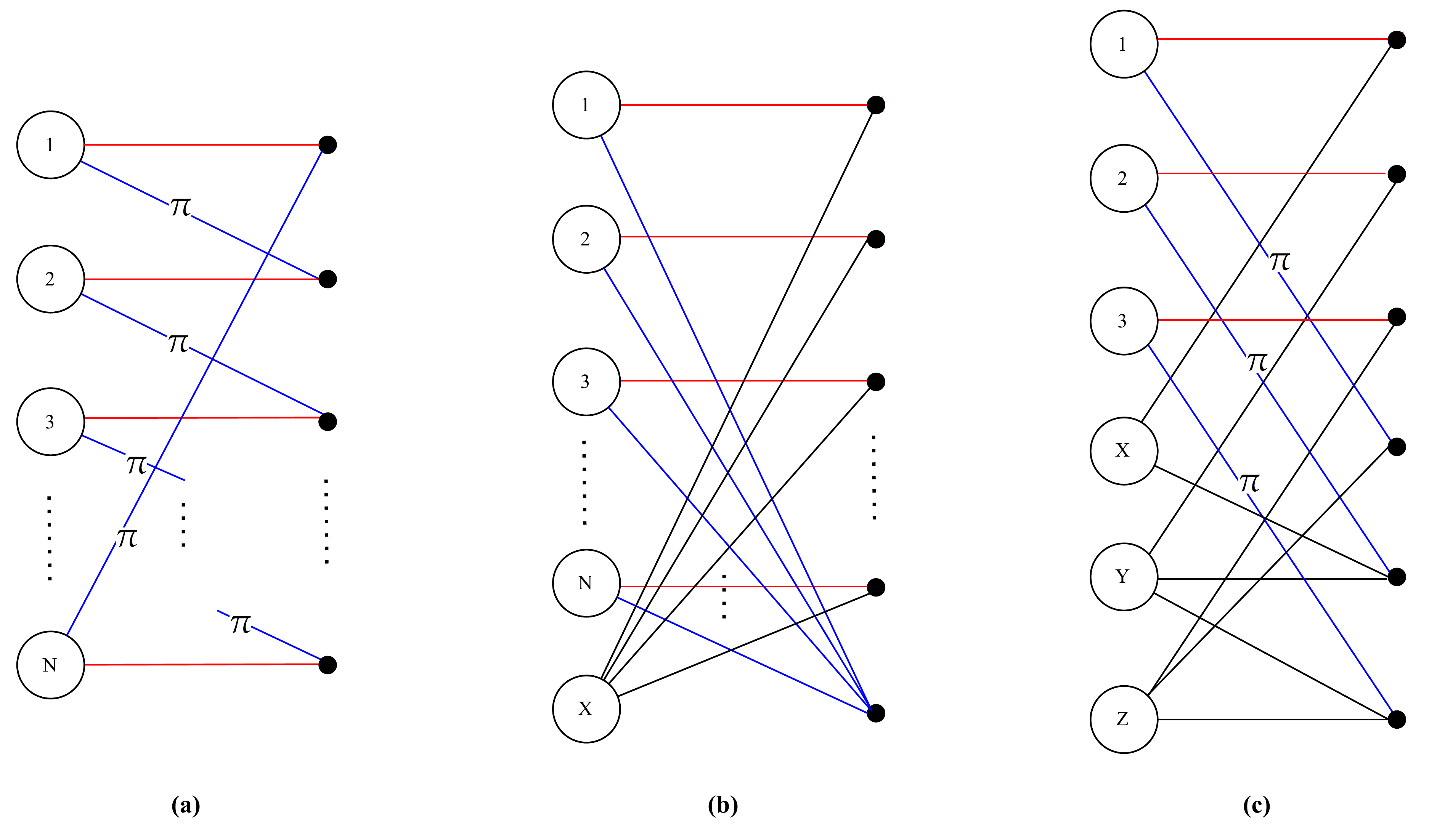}
 \end{gathered}
\end{align}

\section{GHZ state generation scheme}\label{GHZ_cal}

The GHZ generating EPM bigraph (Fig.~1 (a) of the main content) leads to the following operational form
\begin{align}
\hat{A}_{GHZ}|Sym_N\> =
    \frac{1}{\sqrt{2^N}}\prod_{j=1}^N (\ha_{j,+}-\ha_{j\oplus_N 1,-})|Sym_N\> = 
    \frac{1}{\sqrt{2^N}} \Big(\prod_{j=1}^N\ha^\dagger_{j,+} +\prod_{j=1}^N\ha^\dagger_{j,-} \Big)|vac\>,
\end{align} which can be directly checked with Table~1.

Following the translation rules in Fig.~2 of the main concent, we can draw a linear optical GHZ state scheme as follows:
\begin{align}
 \includegraphics[width=10cm]{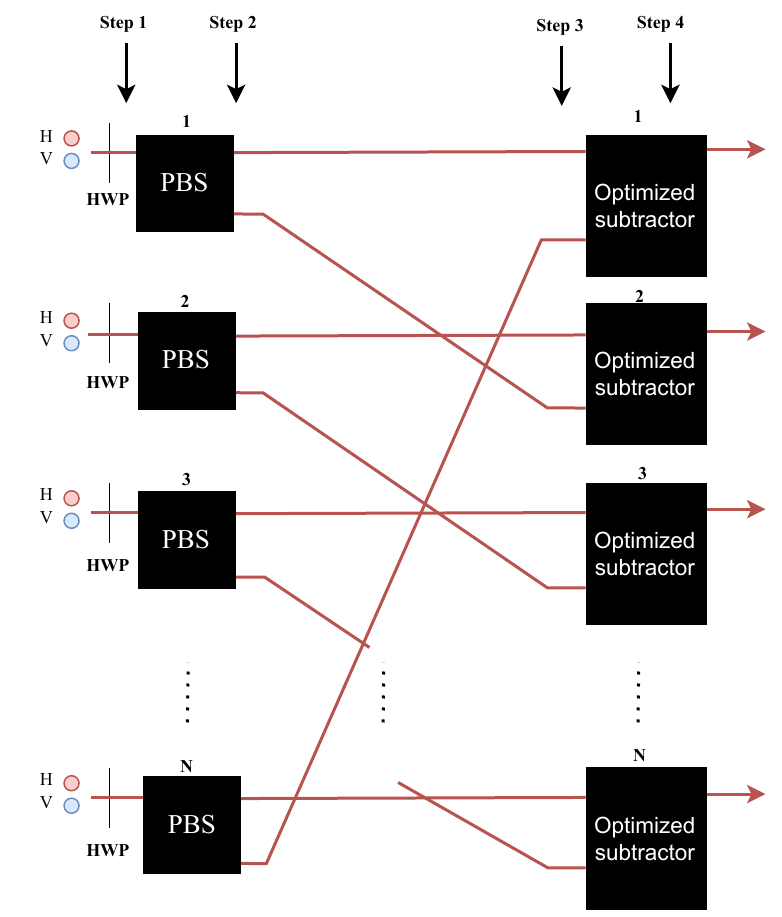}    \nn 
\end{align}

Here spatial modes are labelled as $\{1,2,\cdots, N\}$ in the above circuit. Denoting the creation operator  $\ha^\dagger_{j,s}$ denots that the photon is in spatial mode $j$ ($\in \{1,2,\cdots, N\}$) and polarization $s$ ($\in \{H,V,D,A\}$).

$ $\\
\textbf{Step-by-step explanation}
\begin{itemize}
\item Step 1. Preparation of the initial state
\begin{align}
\prod^{N}_{j=1}\ha^\dagger_{j,H}\ha^\dagger_{j,V}~\to~\prod^{N}_{j=1} \ha^\dagger_{j,D}\ha^\dagger_{j,A} 
\end{align}
\item Step 2.  Divide the photon paths according to the polarization
\begin{align}
\prod^{N}_{j=1} \ha^\dagger_{j,D}\ha^\dagger_{j,A}~\to~
\frac{1}{2^{N}}\prod^{N}_{j=1} (\ha^{\dagger 2}_{j1,H} -\ha^{\dagger 2}_{j2,V})
\end{align}
\item Step 3. Swapping among wires
\begin{align}
\frac{1}{2^{N}}\prod^{N}_{j=1} (\ha^{\dagger 2}_{j1,H} -\ha^{\dagger 2}_{j2,V}) ~\to~
\frac{1}{2^{N}}\prod^{N}_{j=1} (\ha^{\dagger 2}_{j1,H} -\ha^{\dagger 2}_{(j\oplus_{N}1)2,V})
\end{align} where $\oplus_N$ is an addition modulo $N$.
\item Step 4.  Subtractions are performed in two steps. First, the state is transformed by HWP and PBSs in the optimized subtractors as 
\begin{align}
&\frac{1}{2^{N}}\prod^{N}_{j=1} (\ha^{\dagger 2}_{j1,H} -\ha^{\dagger 2}_{(j\oplus_{N}1)2,V})~\to~ \nn \\
&\frac{1}{2^{2N}}\prod^{N}_{j=1}\Big( (\ha^{\dagger}_{j1,H} + \ha^{\dagger}_{j2,V})^2 - (\ha^{\dagger}_{(j\oplus_{N}1)2,H} - \ha^{\dagger}_{(j\oplus_{N}1)1,V})^2\Big) \nn \\
&= 
\frac{1}{2^{2N}}\prod^{N}_{j=1}\Big( (\ha^{\dagger 2}_{j1,H} + 2\ha^{\dagger}_{j1,H} \ha^{\dagger}_{j2,V} +\ha^{\dagger 2}_{j2,V}) - (\ha^{\dagger 2}_{(j\oplus_{N}1)2,H} -2\ha^{\dagger}_{(j\oplus_{N}1)2,H}\ha^{\dagger}_{(j\oplus_{N}1)1,V}  +  \ha^{\dagger2}_{(j\oplus_{N}1)1,V})\Big).
\end{align}
Then we postselect the states with exactly one photon between two detectors attached to the PBS in each optimized subtractor, hence the only states that contribute to the final state are
\begin{align}\label{GHZ_final}
\frac{1}{2^{N-\frac{1}{2}}}\frac{\Big( \prod^{N}_{j=1}(\ha^{\dagger}_{j1,H} \ha^{\dagger}_{j2,V})  
+ \prod^{N}_{j=1}(\ha^{\dagger}_{(j\oplus_{N}1)2,H}\ha^{\dagger}_{(j\oplus_{N}1)1,V})  
\Big)}{\sqrt{2}},
\end{align}
which becomes an $N$-partite GHZ state after the measurements in the optimized subtractors. For example, when all the heralding particles are detected at the uppper mode of the PBSs in the optimized subtractors, the state becomes
\begin{align}
    &\frac{1}{\sqrt{2}}(\prod^N_{j=1} \ha^\dagger_{j1,H} + \prod^N_{j=1}\ha^\dagger_{(j\oplus_{N}1)1,V})|vac\> \nn \\
    &= \frac{1}{\sqrt{2}}(\prod^N_{j=1} \ha^\dagger_{j1,H} +  \prod^N_{j=1}\ha^\dagger_{j1,V})|vac\>.
\end{align}
\end{itemize}
The success probability is determined by the normalization factor of the final state~\eqref{GHZ_final} after the postselection.
The success probability is $\frac{1}{2^{2N}}$ without feed-forward, which becomes   $\frac{1}{2^{2N-1}}$ with feed-forward. 

\newpage
\section{W state  generation scheme}\label{W_cal}

The W generating EPM bigraph (Fig.~\ref{fig_sculpting_bigraphs}, (b)) leads to the following operational form
\begin{align}
\hat{A}_{W}|Sym_{N}\>|Anc_1\> 
&=  
\frac{1}{\sqrt{2^N N}} \Big(\prod_{j=1}^N(\ha_{j+} 
		+ \ha_{X0})\Big)\sum_{k=1}^N\ha_{k-} \Big( \prod_{m=1}^N\ha^\dagger_{m0}\ha^\dagger_{m1} \Big)\ha^\dagger_{X0}|vac\> \nn \\
&= \frac{1}{\sqrt{2^N N}}( \ha^\dagger_{1-}\ha^\dagger_{2+}\cdots \ha^\dagger_{N+} +\ha^\dagger_{1+}\ha^\dagger_{2-}\cdots \ha^\dagger_{N+} +\cdots + \ha^\dagger_{1+}\ha^\dagger_{2+}\cdots \ha^\dagger_{N-})|vac\>.  
\end{align}

Translation rules in Fig.~2 of the main content gives the following circuit:
\begin{align}
 \includegraphics[width=12cm]{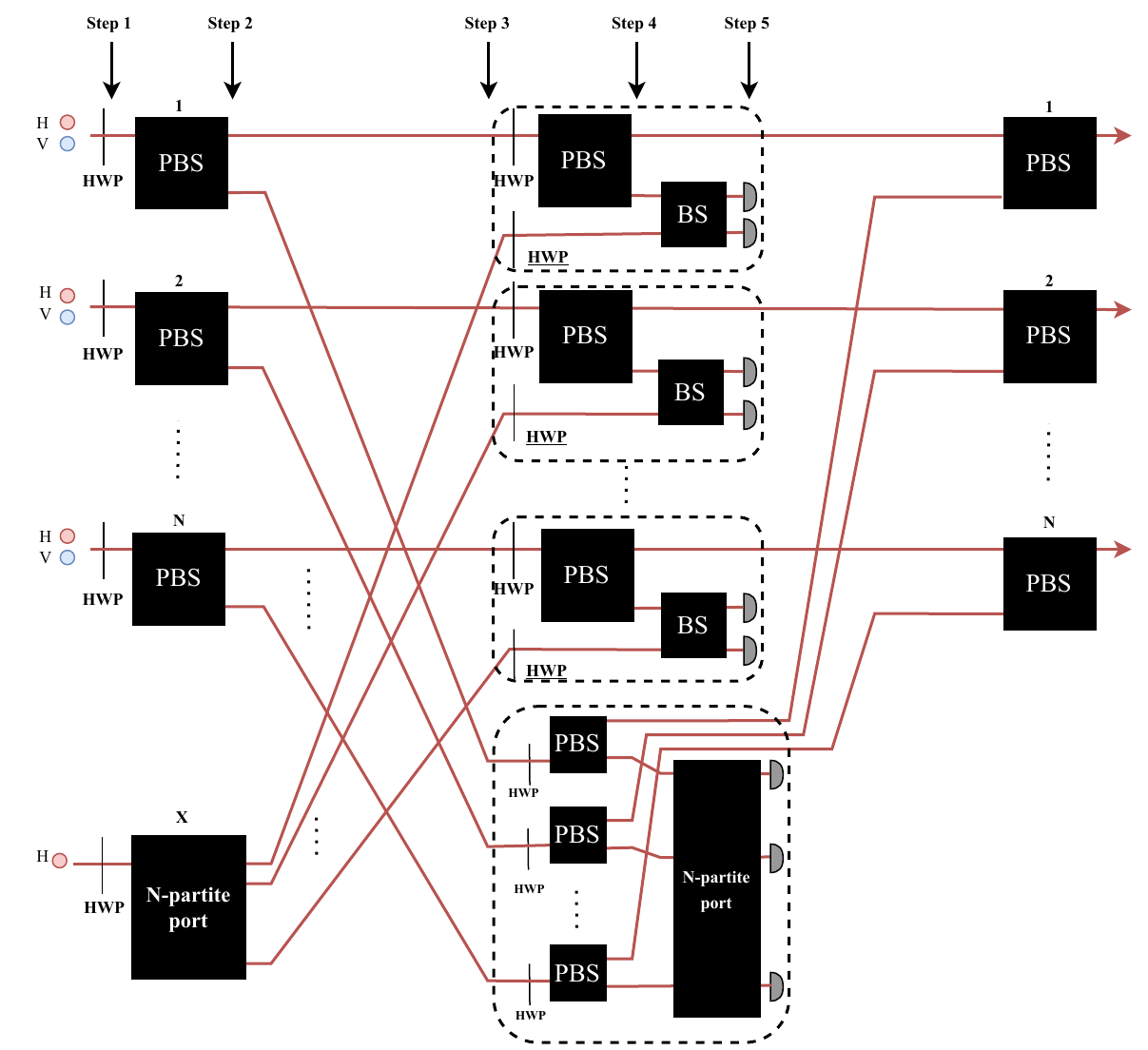}    \nn 
\end{align}
Even if the above circuit is obtained directly from the translation rules from EPM bigaphs to circuits, we can obtain the same final state with fewer operators by changing the initial polarization of the photon in the ancillary mode to $V$ instead of $H$ and removing the HWP before the $N$-partite port and \underline{HWP}s in the subtractors (see Fig.~4 of the main content). Spatial modes are labelled as $\{1,2,\cdots, N, X\}$.

$ $\\
\textbf{Step-by-step explanation}
\begin{itemize}
\item Step 1. Preparation of the initial state
\begin{align}
\Big(\prod^{N}_{j=1}\ha^\dagger_{j,H}\ha^\dagger_{j,V}\Big)\ha^\dagger_{X,H}  ~\to~\Big(\prod^{N}_{j=1} \ha^\dagger_{j,D}\ha^\dagger_{j,A}\Big) \ha^\dagger_{X,D}  
\end{align}
\item Step 2.  Division of the photon paths with $N$ PBSs and one $N$-partite port 
\begin{align}
\Big(\prod^{N}_{j=1} \ha^\dagger_{j,D}\ha^\dagger_{j,A}\Big) \ha^\dagger_{X,D}  ~\to~\frac{1}{2^{N}\sqrt{N}}\prod^{N}_{j=1} (\ha^{\dagger 2}_{j1,H} -\ha^{\dagger 2}_{j2,V})\sum_{k=1}^N\ha^\dagger_{Xk,D}
\end{align}
\item Step 3. Permutation of wires
\begin{align}
\frac{1}{2^{N}\sqrt{N}}\prod^{N}_{j=1} (\ha^{\dagger 2}_{j1,H} -\ha^{\dagger 2}_{j2,V})\sum_{k=1}^N\ha^\dagger_{Xk,D}~\to~\frac{1}{2^{N}\sqrt{N}}\prod^{N}_{j=1} (\ha^{\dagger 2}_{j1,H} -\ha^{\dagger 2}_{Xj,V})\sum_{k=1}^N\ha^\dagger_{k3,D}
\end{align} where we denote $k3$ ($k\in \{1,2,\cdots, N\}$) as the paths from the ancillary $N$-partite port to each detector of the main system.  
\item Step 4. Splitting two-photon states with HWPs and PBSs
\begin{align}
&\frac{1}{2^{N}\sqrt{N}}\prod^{N}_{j=1} (\ha^{\dagger 2}_{j1,H} -\ha^{\dagger 2}_{Xj,V})\sum_{k=1}^N\ha^\dagger_{k3,D} \nn \\
&\to~
\frac{1}{2^{N}\sqrt{N}}\prod^{N}_{j=1} (\ha^{\dagger 2}_{j1,D} -\ha^{\dagger 2}_{Xj,A})\sum_{k=1}^N\ha^\dagger_{k3,V}\nn \\
&\to 
\frac{1}{2^{2N}\sqrt{N}}\prod^{N}_{j=1} \Big( (\ha^{\dagger }_{j1,H} + \ha^{\dagger }_{j2,V} )^2 -(\ha^{\dagger }_{Xj2,H} - \ha^{\dagger }_{Xj1,V} )^2 \Big)\sum_{k=1}^N\ha^\dagger_{k3,V}  \nn \\
&= \frac{1}{2^{2N}\sqrt{N}}\prod^{N}_{j=1} \Big( (\ha^{\dagger 2}_{j1,H} +2\ha^{\dagger }_{j1,H} \ha^{\dagger }_{j2,V}  + \ha^{\dagger 2}_{j2,V} ) -(\ha^{\dagger 2}_{Xj2,H} -2\ha^{\dagger }_{Xj2,H} \ha^{\dagger }_{Xj1,V}  +\ha^{\dagger 2}_{Xj1,V} ) \Big)\sum_{k=1}^N\ha^\dagger_{k3,V}
\end{align}
\item Step 5. Postselection of states with one photon among the detectors attached to the last set of PBSs and $N$-partite port, by which  the states that contribute to the final state are
\begin{align}\label{W_final}
\frac{1}{2^{N}\sqrt{N}}\sum_{k=1}^N
\Big(\ha^\dagger_{k2,V} \ha^{\dagger}_{Xk2,D} \ha^{\dagger }_{k,V}  \prod^{N}_{j\neq k} \ha^{\dagger }_{j,H} \ha^{\dagger }_{j2,V}\Big)
\end{align}

 which becomes the $N$-partite W state with heralding. For example, if all the photons are detected in the upper-most mode in each subtractor, the state becomes
\begin{align}
\frac{1}{\sqrt{N}}\sum_{k=1}^N
\Big(\ha^{\dagger }_{k,V}  \prod^{N}_{j\neq k} \ha^{\dagger }_{j,H}\Big)|vac\>.    
\end{align}
\end{itemize}
The success probability is $\frac{1}{N 2^{2N+1}}$  without feed-forward, which becomes 
$\frac{1}{2^{2N}}$ with feed-forward.

\newpage
\section{$N=3$ Type 5 entangled state  generation scheme}\label{N3_cal}
The $N=3$ type-5 EPM bigraph (Fig.~\ref{fig_sculpting_bigraphs}, (c)) is expressed in the operational form as 
\begingroup
\allowdisplaybreaks
\begin{align}
 \hat{A}_{Type-5}|Sym_{3}\>|Anc_3\> 
&=\frac{1}{12} (\ha_{1+} +\ha_{X0})(\ha_{2+} +\ha_{Y0}) (\ha_{3+} +\ha_{Z0}) \nn \\
&\times (\ha_{Z0} - \ha_{1-})(\ha_{X0} +\ha_{Y0} - \ha_{2-})(\ha_{Y0}+\ha_{Z0} - \ha_{3-}) \Big(\prod_{m=1}^3\ha^\dagger_{m0}\ha^\dagger_{m1}\Big)\ha^\dagger_{X0}\ha^\dagger_{Y0}\ha^\dagger_{Z0}|vac\> \nn \\
&= \frac{1}{12}(\ha^\dagger_{1+}\ha^\dagger_{2+}\ha^\dagger_{3+} + \ha^\dagger_{1-}\ha^\dagger_{2+}\ha^\dagger_{3+} +\ha^\dagger_{1-}\ha^\dagger_{2+}\ha^\dagger_{3-}
+\ha^\dagger_{1-}\ha^\dagger_{2-}\ha^\dagger_{3+} +\ha^\dagger_{1-}\ha^\dagger_{2-}\ha^\dagger_{3-})|vac\> \nn \\
&\sim |+++\> + |-++\> + |-+-\> + |--+\> + |---\>. 
\end{align}
\endgroup
\begin{align}
\includegraphics[width=15cm]{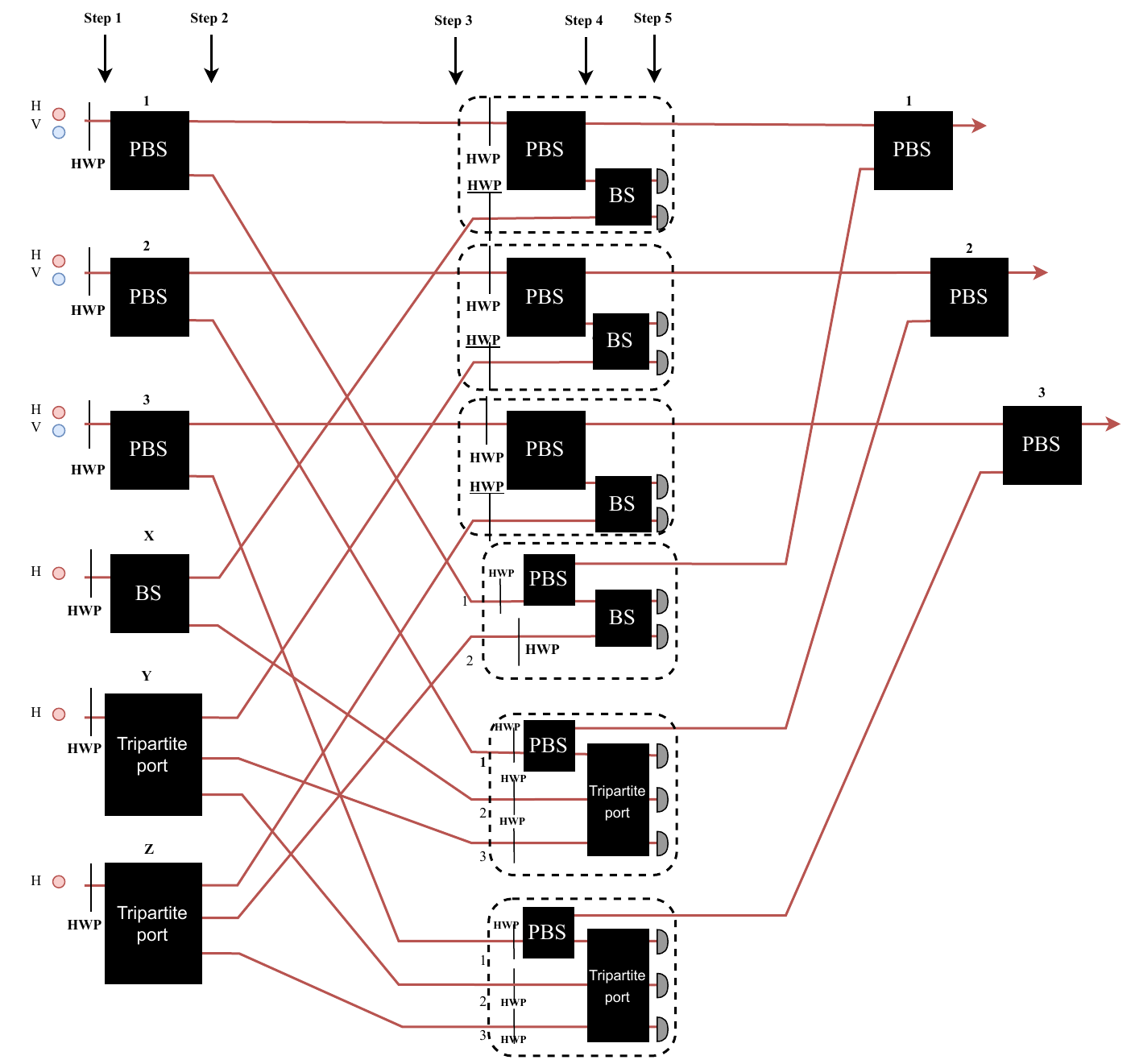} \nn 
\end{align}
 Here spatial modes are denoted as $\{1,2,3, X,Y,Z\}$.

$ $\\
\textbf{Step-by-step explanation}
\begin{itemize}
\item Step 1. Preparation of the initial state
    \begin{align}
\Big(\prod^{3}_{j=1}\ha^\dagger_{j,H}\ha^\dagger_{j,V}\Big)\ha^\dagger_{X,H}  \ha^\dagger_{Y,H} \ha^\dagger_{Z,H} ~\to~\Big(\prod^{N}_{j=1} \ha^\dagger_{j,D}\ha^\dagger_{j,A}\Big) \ha^\dagger_{X,D}  \ha^\dagger_{Y,D} \ha^\dagger_{Z,D}    
    \end{align}
\item Step 2. Division of the photon paths with three PBSs, one BS, and two tripartite ports (tritters)
\begin{align}
\Big(\prod^{N}_{j=1} \ha^\dagger_{j,D}\ha^\dagger_{j,A}\Big) \ha^\dagger_{X,D}  \ha^\dagger_{Y,D} \ha^\dagger_{Z,D} ~\to~\frac{1}{24\sqrt{2}}\prod^{3}_{j=1} (\ha^{\dagger 2}_{j1,H} -\ha^{\dagger 2}_{j2,V})(\ha^\dagger_{X1,D} + \ha^\dagger_{X2,D})\sum_{k=1}^3\ha^\dagger_{Yk,D} \sum_{l=1}^3\ha^\dagger_{Zl,D}
\end{align} Here  ($X,Y,Z$) has (2,3,3) spatial modes respectively, which are labelled as in the three lower purple boxes. 
\item Step 3. Permutation of wires
\begin{align}
 &\frac{1}{24\sqrt{2}} (\ha^{\dagger 2}_{11,H} -\ha^{\dagger 2}_{X1,V})(\ha^{\dagger 2}_{21,H} -\ha^{\dagger 2}_{Y1,V}) (\ha^{\dagger 2}_{31,H} -\ha^{\dagger 2}_{Z1,V}) \nn \\
 &~~~~\times(\ha^\dagger_{13,D} + \ha^\dagger_{Y2,D})(\ha^\dagger_{23,D} + \ha^\dagger_{Y3,D} 
 +\ha^\dagger_{Z2,D}) (\ha^\dagger_{33,D} +\ha^\dagger_{X2,D}+\ha^\dagger_{Z3,D})
\end{align}
where we denote $k3$ ($k\in \{1,2,3\}$) as the paths from the ancillary modes to each detector of the main system.
\item Step 4. Splitting two-photon states with HWPs and PBSs
\begin{align}
 &\frac{1}{24\sqrt{2}} \Big(\ha^{\dagger 2}_{11,D} -\ha^{\dagger 2}_{X1,A}\Big)\Big(\ha^{\dagger 2}_{21,D} -\ha^{\dagger 2}_{Y1,A}\Big) \Big(\ha^{\dagger 2}_{31,D} -\ha^{\dagger 2}_{Z1,A}\Big) \nn \\
 &~~~~\times(\ha^\dagger_{13,V} + \ha^\dagger_{Y2,H})(\ha^\dagger_{23,V} + \ha^\dagger_{Y3,H} 
 +\ha^\dagger_{Z2,H}) (\ha^\dagger_{33,V} +\ha^\dagger_{X2,H}+\ha^\dagger_{Z3,H}) \nn \\
&\to  
\frac{1}{192\sqrt{2}} \Big( (\ha^{\dagger}_{11,H} +\ha^{\dagger}_{12,V})^2 -(\ha^{\dagger}_{X1,H} - \ha^{\dagger}_{1,V})^2 \Big)
\Big( (\ha^{\dagger}_{21,H}+\ha^{\dagger}_{22,V})^2 -(\ha^{\dagger}_{Y1,H} - \ha^{\dagger}_{2,V})^2\Big) \nn \\
&~~~~\times 
\Big( (\ha^{\dagger}_{31,H}+ \ha^{\dagger}_{32,V})^2 -(\ha^{\dagger 2}_{Z1,H} - \ha^{\dagger 2}_{3,V})^2\Big) 
(\ha^\dagger_{13,V} + \ha^\dagger_{Y2,H})(\ha^\dagger_{23,V} + \ha^\dagger_{Y3,H} 
 +\ha^\dagger_{Z2,H}) (\ha^\dagger_{33,V} +\ha^\dagger_{X2,H}+\ha^\dagger_{Z3,H}) 
\end{align}
\item Step 5. Postselection on states with exactly one photon in the sets of detectors belonging to each of the
subtractors, hence the states that contribute to the final states are
\begin{align} 
&\frac{1}{24\sqrt{2} }\Big(
\ha^\dagger_{11,H}\ha^\dagger_{12,V} \ha^\dagger_{21,H}\ha^\dagger_{22,V} \ha^\dagger_{31,H}\ha^\dagger_{32,V}\ha^\dagger_{X2,H}\ha^\dagger_{Y2,H}\ha^\dagger_{Z3,H} 
+\ha^\dagger_{X1,H}\ha^\dagger_{1,V}\ha^\dagger_{21,H}\ha^\dagger_{22,V} \ha^\dagger_{31,H}\ha^\dagger_{32,V}\ha^\dagger_{13,V}\ha^\dagger_{Y2,H}\ha^\dagger_{Z3,H}  \nn \\
&~~~~~~+\ha^\dagger_{X1,H}\ha^\dagger_{1,V}\ha^\dagger_{21,H}\ha^\dagger_{22,V} \ha^\dagger_{Z1,H}\ha^\dagger_{3,V}\ha^\dagger_{13,V}\ha^\dagger_{Y3,H}\ha^\dagger_{33,V}
+ \ha^\dagger_{X1,H}\ha^\dagger_{1,V}\ha^{\dagger}_{Y1,H}\ha^{\dagger}_{2,V}\ha^\dagger_{31,H}\ha^\dagger_{32,V}\ha^\dagger_{13,V}\ha^\dagger_{23,V}\ha^\dagger_{Z3,H}  \nn \\
&
~~~~~+\ha^\dagger_{X1,H}\ha^\dagger_{1,V}\ha^{\dagger}_{Y1,H}\ha^{\dagger}_{2,V}\ha^\dagger_{Z1,H}\ha^\dagger_{3,V}\ha^\dagger_{13,V}\ha^\dagger_{23,V}\ha^\dagger_{33,V} \Big)
\end{align}
For example, if we postselect the cases when a photon arrives at the uppper-most detector in each subtractor, the final state becomes
\begin{align}
 &\frac{1}{\sqrt{5} }\Big(
\ha^\dagger_{11,H} \ha^\dagger_{21,H} \ha^\dagger_{31,H} 
+\ha^\dagger_{1,V}\ha^\dagger_{21,H}\ha^\dagger_{31,H} +\ha^\dagger_{1,V}\ha^\dagger_{21,H} \ha^\dagger_{3,V}
+ \ha^\dagger_{1,V}\ha^{\dagger 2}_{2,V}\ha^\dagger_{31,H}+\ha^\dagger_{1,V}\ha^{\dagger 2}_{2,V}\ha^\dagger_{3,V} \Big)|vac\> 
\end{align}
\end{itemize}
The success probability is $5/(3^2 2^7)$ with feed-forward.

\newpage

\section{From the polarization encoding to the dual-rail encoding}\label{polarization_to_dual}

We can consider the dual-rail encoding counterparts of PBS optical heralded circuits given in Sec.3 to 5. In the dual-rail encoding, two paths in the same spatial mode correspond to the internal state $0$ and $1$. We denote the $j$th spatial mode as
$\begin{gathered}
    \includegraphics[width=2.5cm]{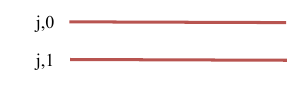}
\end{gathered}$. Then we can obtain dual-rail encoding linear optical schemes by replacing all the PBS operators with dual-rail operators as follows: 
\begin{align}\label{pbs_to_dual}
    \includegraphics[width=.7\textwidth]{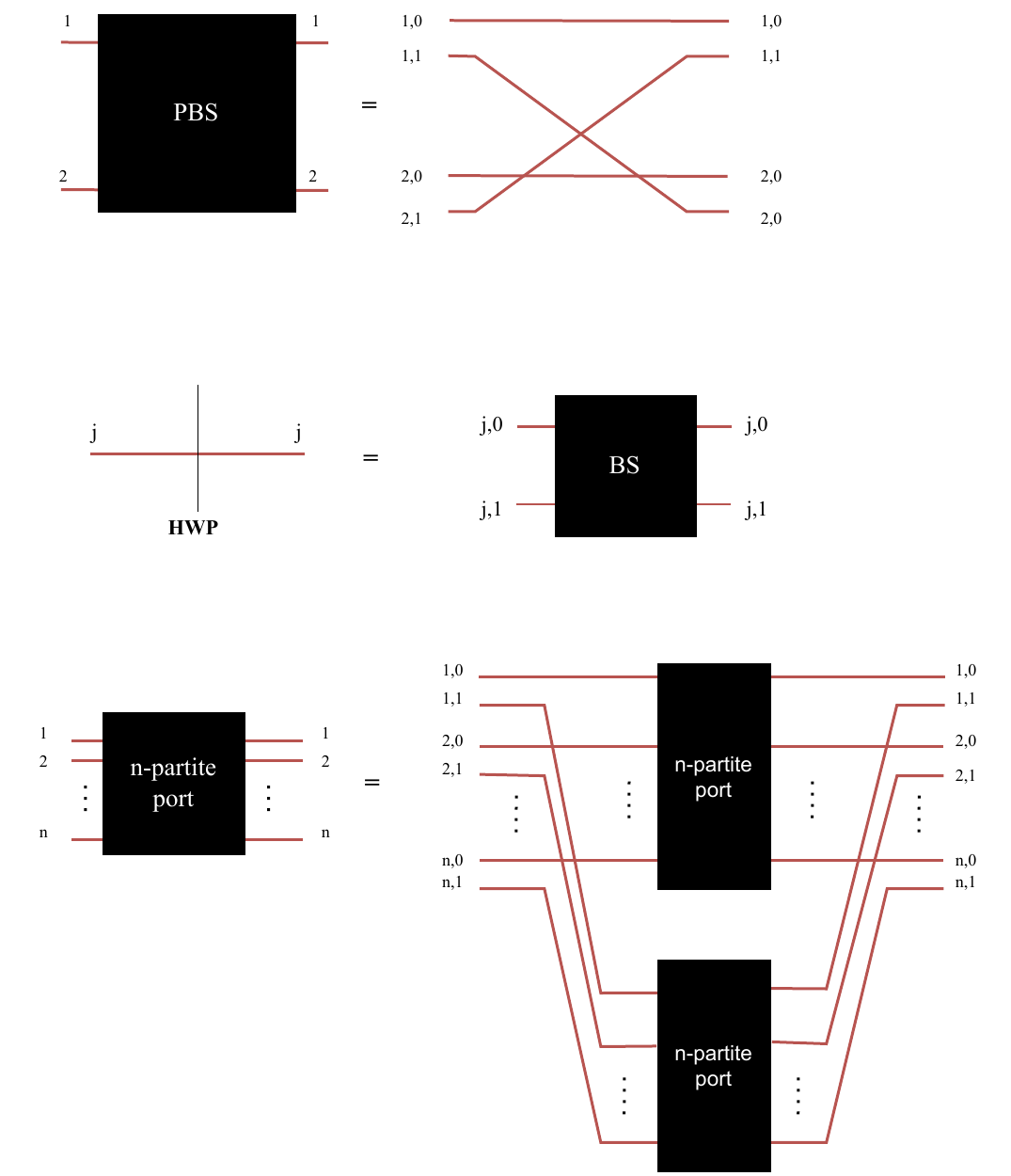} 
\end{align}
In the above correspondence, it is interesting to observe that permutations of wires and BSs in the dual-rail encoding replaces PBSs and HWPs in the polarized encoding respectively.

For example, we can design an $N=3$ W-state generation scheme in the dual-rail encoding with the above correspondence relations~\eqref{pbs_to_dual} as
\begin{align}
    \includegraphics[width=10cm]{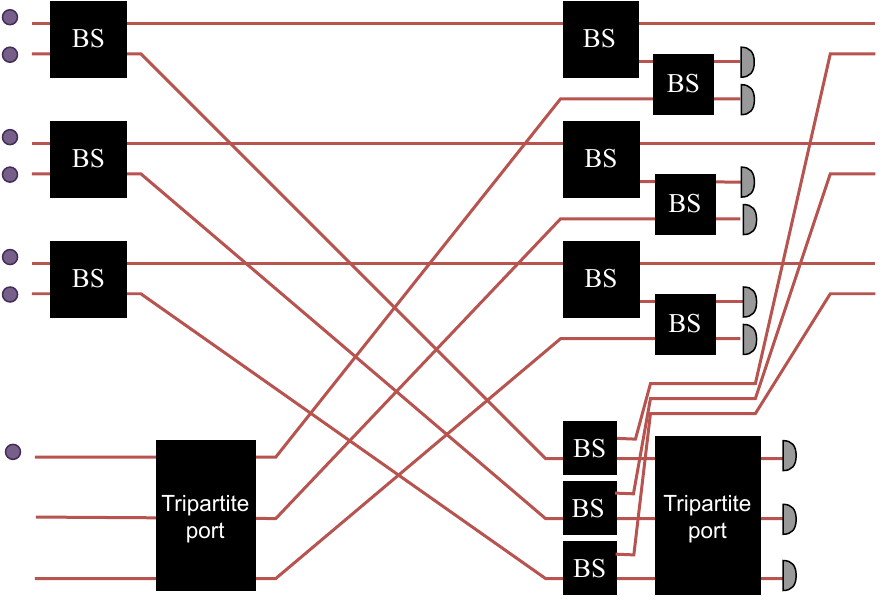} \nn 
\end{align}
Note that the role of PBSs are merged into the permutation of wires for the spatial overlap of photons in the dual-rail version of the scheme.

\end{document}